\documentclass[11pt]{article}
\usepackage{amsmath,amssymb,amsfonts,amsthm}
\usepackage[titletoc,title]{appendix}
\usepackage[english]{babel}
\usepackage[cp1251]{inputenc}
\newcommand{\be}{\begin{equation}}
\newcommand{\ee}{\end{equation}}
\newcommand{\bea}{\begin{eqnarray}}
\newcommand{\eea}{\end{eqnarray}}
\newcommand{\nn}{\nonumber \\}
\newcommand{\p}[1]{(\ref{#1})}
\newcommand{\lb}{\label}

\usepackage[left=1in,right=1in,top=1in,bottom=1in,bindingoffset=0cm]{geometry}
\numberwithin{equation}{section}

\begin{document}
\begin{titlepage}

\hfill{ITP-UH-19/16, {\,} JINR E2-2016-59}

\vfill \vfill \vfill
\begin{center}
{\bf\LARGE SU(2$|$2) SUPERSYMMETRIC MECHANICS}
\end{center}
\vspace{1.5cm}

\begin{center}
{\bf\Large Evgeny Ivanov${\,}^{a)}$, Olaf Lechtenfeld${\,}^{b)}$, Stepan Sidorov${\,}^{a)}$}\\
\end{center}
\vspace{0.4cm}

\centerline{${\,}^{a)}$\it Joint Institute for Nuclear Research,
Dubna, Moscow Region, 141980 Russia}
\vspace{0.2cm}

\centerline{${\,}^{b)}$ \it Institut f\"ur Theoretische Physik
and Riemann Center for Geometry and Physics,}
\centerline
{\it Leibniz Universit\"at Hannover,
Appelstra{\ss}e 2, 30167 Hannover, Germany.}
\vspace{0.3cm}

\centerline{eivanov@theor.jinr.ru,~lechtenf@itp.uni-hannover.de,~sidorovstepan88@gmail.com}
\vspace{0.2cm}

\vspace{2cm}

\par
\begin{center}
{\bf ABSTRACT}
\end{center}

\noindent We introduce a new kind of non-relativistic ${\cal
N}{=}\,8$ supersymmetric mechanics, associated with worldline
realizations of the supergroup $SU(2|2)$ treated as a deformation of flat ${\cal
N}{=}\,8$, $d{=}1$ supersymmetry. Various worldline $SU(2|2)$
superspaces are constructed as coset manifolds of this supergroup,
and the corresponding superfield techniques are developed. For the
off-shell $SU(2|2)$ multiplets $({\bf 3,8,5})$, $({\bf 4,8,4})$ and
$({\bf 5,8,3})$, we construct and analyze the most general
superfield and component actions. Common features are mass
oscillator-type terms proportional to the deformation parameter and
a trigonometric realization of the superconformal group $OSp(4^*|4)$
in the conformal cases. For the simplest $({\bf 5, 8, 3})$ model the
quantization is performed.

\vspace{5cm}

\end{titlepage}

\setcounter{equation}{0}
\section{Introduction}
In recent years, interest has grown in theories invariant under some ``curved'' analogs
of rigid Poincar\'e supersymmetry in diverse dimensions \cite{FS,DFS,Closset}.
The main motivation was to check general gauge/gravity correspondences in concrete
field-theoretical examples, classically as well as quantum mechanically.
One construction of such theories is by the localization method~\cite{Pestun}, which proceeds
from the relevant supergravity theories in component formulation. Alternatively, one can start
from the supergroup of the corresponding ``curved'' supersymmetry,
list its various coset superspaces and develop appropriate superfield techniques. These permit
the derivation of invariant actions as superspace Berezin integrals, with Lagrangians being
functions of superfields and their covariant derivatives. This second approach
was used in \cite{Kuz,SamSor1,SamSor2} and goes back to~\cite{IvSo} where superfield
techniques for $OSp(1|4)$ supersymmetry in four dimension were fully developed for the first time.

Supersymmetric mechanics~\cite{Witten1,Witten2} represents the extreme $d{=}1$ case
of Poincar\'e-supersymmetric field theory.
In the underlying $d{=}1$ ``Poincar\'e superalgebra'' the supercharges square to the Hamiltonian
(and perhaps some constant or operator-valued central charges).
Mechanical analogs of higher-dimensional curved rigidly supersymmetric theories can be based
on semi-simple supergroups which yield flat $d{=}1$ supersymmetries through some contraction.
In other words, mechanical models on such supergroups can be treated as deformations of
standard supersymmetric mechanics. The main difference between the two types of
supersymmetric mechanics models lies in the closure of the supercharges:
In the deformed case it contains not only the Hamiltonian but also generators of some
nontrivial internal symmetry. As a consequence, the corresponding Hilbert spaces and spectra
essentially differ from each other. In particular, in the deformed case an energy level may carry
unequal numbers of bosonic and fermionic states.

The simplest examples of such deformed supersymmetric mechanics substitute flat
${\cal N}{=}\,4, d{=}1$ supersymmetry with the supergroup $SU(2|1)$. At the component level,
they were constructed in~\cite{WS,BelNer1,BelNer2,R1,R2}.
The $SU(2|1)$ symmetry in the model built on the multiplet $({\bf 1,4,3})$~\footnote{
Multiplets of the standard and deformed ${\cal N}{=}\,4$, $d{=}1$ supersymmetry are denoted as 
${\bf (k,4,4{-}k)}$ with ${\bf k}{=}0,1,2,3,4$. These numbers correspond to the numbers of 
bosonic physical fields, fermionic physical fields and bosonic auxiliary fields, respectively. 
${\cal N}{=}\,8$, $d{=}1$ multiplets are denoted in the same way as ${\bf (k,8,8{-}k)}$, 
where ${\bf k}{=}0,1,\ldots,8$.} 
has been considered first in~\cite{WS} and christened there ``weak supersymmetry''.
Models based on the chiral $({\bf 2,4,2})$ multiplet were considered in~\cite{BelNer1,BelNer2,R1,R2},
including a computation of the superconformal index.

The superfield approach to $SU(2|1)$ mechanics was worked out in~\cite{DSQM,SKO,ISTconf,DHSS}.
The superfield techniques not only reproduced the models of~\cite{WS,BelNer1,BelNer2,R1,R2}
(and their interacting extensions)
but also revealed some new models, in particular those associated with the $({\bf 4,4,0})$ multiplet.
The latter admits a harmonic $SU(2|1)$ superspace description generalizing
flat ${\cal N}{=}\,4$, $d{=}1$ harmonic superspace~\cite{IvLe}.
Many notions and models of ${\cal N}{=}\,4$ mechanics still await a deformation
to the $SU(2|1)$ case, for instance
${\cal N}{=}\,4$ supersymmetric Calogero-Moser systems~\cite{FeIvLe1,FeIvLe2},
the gauging procedure in ${\cal N}{=}\,4$ mechanics~\cite{DeIv1,DeIv2}, or
${\cal N}{=}\,4$ models coupled to background abelian or non-abelian gauge fields
\cite{self1,self2,self3}.

The present paper makes one step in a new direction, initiating a study of deformed supersymmetric
mechanics models associated with worldline realizations of the supergroup $SU(2|2)$.
This supergroup contains eight supercharges and so can be viewed as a deformation
of flat ${\cal N}{=}\,8$, $d{=}1$ supersymmetry.
${\cal N}{=}\,8$ mechanics has appeared in many contexts
(see, e.g.\cite{Diac,BIKL1,IvSmi,BIKL2,286,BHSS,ILS,NewM,KuTo1,KuTo,KhTo,GKLT,Dorey1,Dorey2}).
Our aim here is to construct and discuss $SU(2|2)$ analogs for some of these models,
employing the appropriate worldline superfield approach generalizing the $SU(2|1)$ one.
Our consideration relies essentially on~\cite{BIKL1,BIKL2,BHSS,ILS},
where superfield methods were efficiently applied for flat ${\cal N}{=}\,8$, $d{=}1$ supersymmetry.
In the contraction limit, when $SU(2|2)$ goes over into its flat counterpart,
the models of this paper specialize to those considered in~\cite{BIKL1,BIKL2,BHSS,ILS}.

In \cite{BMN}, Berenstein, Maldacena and Nastase proposed an M-theory matrix model
with $16$ supercharges, which spurred investigations of massive super Yang-Mills mechanics
(see, e.g.~\cite{DSV1,DSV2,KP1,Motl,KP2,Denef}).
Since their matrix model has $SU(2|4)$ supersymmetry~\cite{DSV2,KP1},
$SU(2|1)$ and $SU(2|2)$ supersymmetric mechanics are expected to be relevant for
massive matrix models with $4$ and $8$ supercharges, respectively.

The plan of the paper is as follows.
In Section~2 we describe coset superspaces of $SU(2|2)$, to be used in the following sections
for defining superfields carrying various irreducible $SU(2|2)$ multiplets.
Besides the standard real $SU(2|2)$ superspace we introduce the chiral superspace,
the harmonic superspace and the biharmonic superspace.
We define the necessary elements of the corresponding superfield technique:
covariant derivatives, transformation laws, and invariant integration measures.
In Sections 3, 4 and~5 we present the models associated with the off-shell $SU(2|2)$ multiplets
$({\bf 3,8,5})$, $({\bf 4,8,4})$ and $({\bf 5,8,3})$.
We give both the superfield and component-field actions for all cases.
Some of these actions reveal enhanced superconformal-type symmetries, some do not.
Common features of most actions are an oscillator-type mass term for the fields
and a trigonometric realization of the superconformal symmetries.
As an example of a quantum model, in Subsection~5.3 we discuss $SU(2|2)$ quantum mechanics
based on a free $({\bf 5,8,3})$ multiplet.
In the concluding Section~6 we mention links with other models and outline some directions
for further study. We also adduce arguments why certain flat ${\cal N}{=}\,8, d{=}1$ multiplets
(in particular the ``root'' multiplet $({\bf 8,8,0})$) seem not to admit a deformation
to $SU(2|2)$ mechanics. We transferred into three Appendices some technical points,
including the calculation of various harmonic integrals,
the embedding of the superalgebra $su(2|2)$
into the ${\cal N}{=}\,8, d{=}1$ superconformal algebra $osp(4^*|4)$,
the realization of the latter on the $SU(2|2)$ multiplets considered,
as well as a short account of the off-shell $SU(2|1)$ multiplet $({\bf 3,4,1})$.
The latter is an important constituent of our $SU(2|2)$ multiplets
but was not properly treated in previous papers on $SU(2|1)$ mechanics.

\setcounter{equation}{0}
\section{Deformed ${\cal N}{=}\,8$, $d{=}1$ superspaces}
In this section, we formulate a deformed real ${\cal N}{=}\,8$, $d{=}1$ superspace where worldline realizations of the supergroup $SU(2|2)$ can be defined.
Then we construct the corresponding chiral, analytic harmonic and analytic biharmonic $SU(2|2)$ superspaces. In the following sections,
these types of superspaces will be used for defining different types of superfields and for setting up $SU(2|2)$ invariant actions of the latter, generalizing
those constructed in \cite{BIKL1,BIKL2,BHSS,ILS} in the presence of flat ${\cal N}{=}\,8$, $d{=}1$ supersymmetry.

\subsection{Superalgebra}
Our starting point is the superalgebra $su(2|2)$ with three central charges:
\bea
    &&\left\lbrace Q^{ia}, S^{jb}\right\rbrace = 2im\left(\varepsilon^{ab}L^{ij} -  \varepsilon^{ij}R^{ab}\right) + 2\,\varepsilon^{ab}\varepsilon^{ij}C,\nn
    &&\left\lbrace Q^{ia}, Q^{jb}\right\rbrace = 2\,\varepsilon^{ij}\varepsilon^{ab}
    \left(H+C_1\right),\qquad \left\lbrace S^{ia}, S^{jb}\right\rbrace = 2\,\varepsilon^{ij}\varepsilon^{ab}\left(H-C_1\right),\nn
    &&\left[L^{ij}, L^{kl}\right] = \varepsilon^{il}L^{kj} +\varepsilon^{jk}L^{il}, \qquad \left[R^{ab}, R^{cd}\right] = \varepsilon^{ad}R^{bc} +\varepsilon^{bc}R^{ad},\nn
    && \left[L^{ij}, Q^{ka}\right] = \frac{1}{2}\left(\varepsilon^{ik}Q^{ja} + \varepsilon^{jk}Q^{ia}\right), \qquad
    \left[R^{ab}, Q^{kc}\right] = \frac{1}{2}\left(\varepsilon^{ac}Q^{kb} + \varepsilon^{bc}Q^{ka}\right),\nn
    && \left[L^{ij}, S^{ka}\right] = \frac{1}{2}\left(\varepsilon^{ik}S^{ja} + \varepsilon^{jk}S^{ia}\right), \qquad
    \left[R^{ab}, S^{kc}\right] = \frac{1}{2}\left(\varepsilon^{ac}S^{kb} + \varepsilon^{bc}S^{ka}\right).
\label{algebra}
\eea
All other (anti)commutators are vanishing.

The superalgebra $su(2|2)$ contains in general three central charges $C$,
$C_1$ and $H$. The generators $L^{ij} = L^{ji}$, $R^{ab} = R^{ba}$
form two mutually commuting $su(2)$ algebras, $su(2)_{\rm L}$ and
$su(2)_{\rm R}$. The conjugation rules are as follows:\footnote{The
doublet indices are raised and lowered in the standard way by the
$\varepsilon$ symbols, e.g.,
$$Q_{ia} = \varepsilon_{ij}\varepsilon_{ab}Q^{jb},\qquad \varepsilon_{12}= -\,\varepsilon^{12} =1\,.$$}
\be
    \left(Q_{ia}\right)^\dagger = Q^{ia},\;\; \left(S_{ia}\right)^\dagger = S^{ia},\;\;
    \left(L_{ij}\right)^\dagger = - L^{ij},\;\; \left(R_{ab}\right)^\dagger = - R^{ab},\;\;     
    H^\dagger = H,\;\;\left(C\right)^\dagger = C,\;\; \left(C_1\right)^\dagger = C_1\,.
\ee
The mass dimension parameter $m$ plays the same role as in the $SU(2|1)$ case: by contraction $m \rightarrow 0$ the relations
\p{algebra} are reduced to those of the flat ${\cal N}{=}\,8, d{=}1$ ``Poincar\'e'' superalgebra extended by central charges $C$, $C_1$ and possessing
a restricted $R$-symmetry group $SO(4) \sim SU(2)_{\rm L}\times SU(2)_{\rm R}${ }.\footnote{
If $C=C_1=0$\,, the $R$-symmetry group enhances to $SO(8)$.} Correspondingly,
\p{algebra} can be considered as a deformation of the flat ${\cal N}{=}\,8, d{=}1$ supersymmetry, with $m$ as a deformation parameter.

To understand the origin of the central charge operators in \p{algebra}, let us note that these relations in fact coincide with those defining
a deformation of the flat ${\cal N}{=}(4,4)$, $d{=}2$ Poincar\'e superalgebra. Indeed, in the $m=0$ limit \p{algebra} can be identified with
a sum of two independent ${\cal N}{=}4\,, d{=}2$ algebras in the left and right sectors of $d{=}2$ Minkowski space-time in the light-cone parametrization,
with $H+C_1$ and $H - C_1$ being the mutually commuting translation operators along two light-cone directions. Moreover, one can realize the $d{=}2$ Lorentz
group $SO(1,1)$ as an additional automorphism group of \p{algebra} acting as real rescalings of the mutually (anti)commuting sets $(Q^{ia}, H + C_1)$
and $(S^{ia}, H - C_1)$ (with the weights $(1/2, 1)$ and $(-1/2, -1)$, respectively). In such an interpretation, the generator $C$ in \p{algebra}
is $SO(1,1)$ singlet and so it is the central charge from the $d{=}2$ perspective as well, while $C_1$ generates the translation along the spatial $d{=}2$ direction.
The natural and simplest reduction from $d{=}2$ to $d{=}1$ proceeds by eliminating altogether the dependence on the spatial coordinate,
i.e. just by putting to zero the generator $C_1$. In what follows we will deal with such a limited $su(2|2)$ superalgebra, corresponding to the choice $C_1=0$ in
\p{algebra}. In principle, it is easy to construct the $SU(2|2), d{=}1$ superfield formalism with $C_1\neq 0$,\footnote{Non-zero $C_1$ could be generated
within the Scherk-Schwarz type $d{=}2 \rightarrow d{=}1$  reduction (see, e.g., \cite{PSBN4}).} but in all examples considered below there is no need to
activate this central charge. It is not the case for the ``genuine'' central charge $C$ which defines an actual symmetry, e.g., in the models based on the multiplet
$({\bf 4, 8, 4})$ (Sect. 4).

One can rewrite the superalgebra \eqref{algebra} (hereafter with $C_1=0$) in a different form by defining the complex supercharges
\bea
    \Pi^{ia}:=\frac{1}{\sqrt{2}}\left(Q^{ia}-iS^{ia}\right),\qquad
    \bar{\Pi}^{ia} := \big(\Pi_{ia}\big)^\dagger = \frac{1}{\sqrt{2}}\left(Q^{ia}+iS^{ia}\right).
\eea
In the complex basis, the (anti)commutators of \eqref{algebra} become
\bea
    &&\left\lbrace \Pi^{ia}, \bar{\Pi}^{jb}\right\rbrace = -\,2m\left(\varepsilon^{ab}L^{ij} -  \varepsilon^{ij}R^{ab}\right) + 2\,\varepsilon^{ab}\varepsilon^{ij}H,\nn
    &&\left\lbrace \Pi^{ia}, \Pi^{jb}\right\rbrace = -\,2i\,\varepsilon^{ij}\varepsilon^{ab}C,\qquad \left\lbrace \bar{\Pi}^{ia}, \bar{\Pi}^{jb}\right\rbrace = 2i\,\varepsilon^{ij}\varepsilon^{ab}C,\nn
    &&\left[L^{ij}, L^{kl}\right] = \varepsilon^{il}L^{kj} +\varepsilon^{jk}L^{il}, \qquad \left[R^{ab}, R^{cd}\right] = \varepsilon^{ad}R^{bc} +\varepsilon^{bc}R^{ad},\nn
    && \left[L^{ij}, \Pi^{ka}\right] = \frac{1}{2}\left(\varepsilon^{ik}\Pi^{ja} + \varepsilon^{jk}\Pi^{ia}\right), \qquad
    \left[R^{ab}, \Pi^{kc}\right] = \frac{1}{2}\left(\varepsilon^{ac}\Pi^{kb} + \varepsilon^{bc}\Pi^{ka}\right),\nn
    && \left[L^{ij}, \bar{\Pi}^{ka}\right] = \frac{1}{2}\left(\varepsilon^{ik}\bar{\Pi}^{ja} + \varepsilon^{jk}\bar{\Pi}^{ia}\right), \qquad
    \left[R^{ab}, \bar{\Pi}^{kc}\right] = \frac{1}{2}\left(\varepsilon^{ac}\bar{\Pi}^{kb} + \varepsilon^{bc}\bar{\Pi}^{ka}\right).\label{algebra1}
\eea

The supergroup $SU(2|2)$ contains a few $SU(2|1)$ subgroups.
One of them has the bosonic subgroup $SU(2)_{\rm L}\times U(1)_{\rm R}$ with $U(1)_{\rm R} \subset SU(2)_{\rm R}$, while another has the bosonic
subgroup $SU(2)_{\rm R}\times U(1)_{\rm L}$ with $U(1)_{\rm L} \subset SU(2)_{\rm L}$.
These supergroups are equivalent up to switching  $SU(2)_{\rm R}\leftrightarrow SU(2)_{\rm L}$.
In what follows, we will mainly deal with the first choice, where
$SU(2|1)$ generators \cite{DSQM}
are singled out as~\footnote{The other pair of supercharges
$\Pi^{i2}$, $\bar{\Pi}_{j2}$ also form an $su(2|1)$ superalgebra, with the same set of bosonic generators.}
\bea
&& \Pi^{i1}=:Q^{i},\quad \bar{\Pi}_{j1}=:\bar{Q}_{j}\,,\quad L^{i}_{j}=:I^i_j\,,\quad R^{12}=:F, \label{leftSU2} \\
&& \{Q^i, \bar Q_j\} = 2\delta^i_j \left(H - mF\right) + 2m I^i_j\,, \; \{Q^i, Q^j \} = 0\,, \;\; \left[F, Q^i\right] =\frac12\, Q^i,\; \left[F, \bar{Q}_i\right] = -\frac12\, \bar Q_i\,. \label{leftSU2ant}
\eea
The second basic $su(2|1)$ subalgebra is formed by the generators
$\Pi^{1a}$, $\bar{\Pi}_{1b}$, $R^{ab}$, $L^{12}$, $H$.
Actually, the generators \eqref{leftSU2} form the {\it centrally extended} superalgebra $\hat{su}(2|1)$ with the central charge $H$,
and the same is true for the second $SU(2|1)$.
The central charge $H$ in \eqref{leftSU2} is a difference of external and internal $U(1)$ generators in the extended superalgebra $su(2|1)\oplus u(1)_{\rm ext}$ \cite{SKO}.
If the generator $H-mR^{12}$ is chosen as the full internal $U(1)$ generator of $su(2|1)$ (such a choice is admissible since $H$ commutes with everything),
then $R^{12}$ decouples and becomes a generator of the external $U(1)_{\rm ext}$
$R$-symmetry, such that $u(1)_{\rm ext} \subset su(2)_{\rm R}$\,.

\subsection{Basic $SU(2|2)$ supercoset, Cartan forms and covariant derivatives}
We will be first interested in the realization of $SU(2|2)$ supersymmetry in a real ${\cal N}{=}\,8$, $d{=}1$ superspace identified with the following
supercoset of the supergroup with the superalgebra \eqref{algebra}:
\bea
    \frac{PSU(2|2)\times\mathbb{R}^2}{SU(2)_{\rm L} \times SU(2)_{\rm R} \times \mathbb{R}^1}=
    \frac{\left\lbrace Q^{ia}, S^{jb},L^{ij},R^{ab},C,H\right\rbrace}{\left\lbrace L^{ij},R^{ab},C\right\rbrace}\,.\label{coset}
\eea
Here, the supergroup $PSU(2|2)$ is a corresponding supergroup $SU(2|2)$ without central charges. Further, we will use the notation $SU(2|2)$ as a supergroup with central extensions.
An element of this supercoset is defined as
\bea
    g = e^{\theta_{ia}Q^{ia}}e^{\hat{\theta}_{ia}S^{ia}}e^{i t H}, \qquad g^\dagger = g^{-1},\label{Coset}
\eea
and the supercoset parameters are treated as a set of superspace coordinates
\bea
    \zeta = \left\lbrace t,\theta^{ia},\hat{\theta}^{jb}\right\rbrace ,\qquad
    \overline{(\theta_{ia})}=\theta^{ia},\quad\overline{(\hat{\theta}_{ia})}=\hat{\theta}^{ia}.\label{sspace}
\eea
The central charge generator $H$ is associated with a translation generator along $\mathbb{R}^2/\mathbb{R}^1 \sim \mathbb{R}^1$ parametrized by the time coordinate $t$.

Before presenting the realization of $SU(2|2)$ on these coordinates induced by the left shifts of the element \p{Coset},
it will be convenient to calculate the left-covariant Cartan 1-forms defined by
\bea
    g^{-1}d g = i\,\Delta t\,H + \Delta \theta_{ia}\,Q^{ia} + \Delta\hat{\theta}_{ia}\,S^{ia}
    + \Omega_{ij}\,L^{ij} + \Omega_{ab}\,R^{ab} + \Omega_{(C)}\,C.
\eea
The explicit expressions for these forms are
\bea
    &&\Delta t=dt - i\theta^{ia} d\theta_{ia} - i\hat{\theta}^{ia}d\hat{\theta}_{ia}
    +\frac{4m}{3}\,\hat{\theta}^{jd}\hat{\theta}^i_d\hat{\theta}^a_{(i}d \theta_{j)a}\,, \nn
    &&\Delta \theta^{ia} = d \theta^{ia}, \qquad \Delta \hat{\theta}^{ia} =
d \hat{\theta}^{ia} + 2im\,\hat{\theta}^{ja}\hat{\theta}^{ib}d\theta_{jb}\,, \nn
    &&\Omega_{ij} = 2im \,\hat{\theta}^a_{(i}d\theta_{j)a}\,, \qquad \Omega_{ab} = 2im \,\hat{\theta}_{k(a} d\theta_{b)}^{k}\,,\qquad
    \Omega_{(C)} =  2\,\hat{\theta}_{ia} d\theta^{ia}.
    \label{cartan1}
\eea

The covariant derivatives of some superfield $\Phi^A\big(t,\theta^{ia}, \hat{\theta}^{jb}\big)$ can be found from the general expression
for its covariant differential
\bea
    \Delta \Phi^A  &=& d \Phi^A + \left[\Omega_{ij}\,\tilde{L}^{ij} + \Omega_{ab} \,\tilde{R}^{ab}+ \Omega_{(C)}\,\tilde{C}\right]^A_{\;\;B}\,\Phi^B \nn
    &=&\left[ \Delta t \,D_{(t)} + \Delta \theta_{ia}D^{ia} + \Delta\hat{\theta}_{ia}\nabla^{ia}\right]\Phi^A. \label{defCovD}
\eea
Here, $\tilde{L}^{ij}$ and $\tilde{R}^{ab}$ are ``matrix'' parts of the full $SU(2)$ generators (realized as well on Grassmann coordinates),
which act on the external indices of covariant derivatives as
\bea
    \tilde{L}^{ij}D^{ka} = -\,\frac{1}{2}\left(\varepsilon^{ik}D^{ja} + \varepsilon^{jk}D^{ia}\right), \qquad
    \tilde{R}^{ab}D^{kc} = -\,\frac{1}{2}\left(\varepsilon^{ac}D^{kb} + \varepsilon^{bc}D^{ka}\right),\qquad {\rm etc}\,.
\eea
In the same way, they act on the external $SU(2)_{\rm L}\times SU(2)_{\rm R}$ indices of superfields.
The rule of complex conjugation for these matrix parts is as follows
\bea
    \big(\tilde{L}_{ij}\big)^\dagger =  \tilde{L}^{ij},\qquad \big(\tilde{R}_{ab}\big)^\dagger = \tilde{R}^{ab},\qquad
    \big(\tilde{C}\,\big)^\dagger = -\,\tilde{C}.
\eea

Explicitly, the covariant derivatives are given by the following expressions
\bea
    D^{ia} &=& \frac{\partial}{\partial \theta_{ia}} - i\left(\theta^{ia}-\frac{2im}{3}\,\hat{\theta}^{ib}\hat{\theta}^{ja}\hat{\theta}_{jb}\right)\partial_{t} - 2\,\hat{\theta}^{ia}\tilde{C} - 2im\,\hat{\theta}^{ib}\hat{\theta}^{ja}\frac{\partial}{\partial \hat{\theta}^{jb}} \nn
    && -\, 2im\left[\hat{\theta}_{j}^{a}\,\tilde{L}^{ij} - \hat{\theta}_{b}^{i}\,\tilde{R}^{ab}\right], \nn
    \nabla^{ia} &=& \frac{\partial}{\partial \hat{\theta}_{ia}} - i\hat{\theta}^{ia}\partial_{t}\,,\nn
    D_{(t)} &=&\partial_{t}=\frac{\partial}{\partial t}\,.\label{CovD}
\eea
They satisfy the anticommutation relations
\bea
    &&\left\lbrace D^{ia},\nabla^{jb}\right\rbrace = -\,2im\left(\varepsilon^{ab} \tilde{L}^{ij} - \varepsilon^{ij}\tilde{R}^{ab} \right) -2\,\varepsilon^{ab}\varepsilon^{ij}\tilde{C},\nn
    &&\left\lbrace D^{ia}, D^{jb}\right\rbrace = -\,2i\,\varepsilon^{ij}\varepsilon^{ab}\partial_{t}\,,\qquad
    \left\lbrace \nabla^{ia}, \nabla^{jb}\right\rbrace = -\,2i\,\varepsilon^{ij}\varepsilon^{ab}\partial_{t}\,.
\eea
By rewriting the covariant derivative as
\bea
    {\cal D}^{ia} = \frac{1}{\sqrt{2}}\left(D^{ia}-i\nabla^{ia}\right),\qquad
    \bar{\cal D}^{ia} = \frac{1}{\sqrt{2}}\left(D^{ia}+i\nabla^{ia}\right),
\eea
we obtain that
\bea
    &&\left\lbrace {\cal D}^{ia},\bar{\cal D}^{jb}\right\rbrace = 2m\left(\varepsilon^{ab} \tilde{L}^{ij} - \varepsilon^{ij}\tilde{R}^{ab} \right)
    - 2i\,\varepsilon^{ab}\varepsilon^{ij}\partial_{t}\,,\label{DbarD} \\
    &&\left\lbrace {\cal D}^{ia}, {\cal D}^{jb}\right\rbrace = 2i\,\varepsilon^{ij}\varepsilon^{ab}\tilde{C},\qquad
    \left\lbrace \bar{\cal D}^{ia}, \bar{\cal D}^{jb}\right\rbrace = -\,2i\,\varepsilon^{ij}\varepsilon^{ab}\tilde{C}. \label{DD}
\eea

\subsection{Transformation properties}
The transformation properties of the ${\cal N}{=}\,8$ superspace coordinates under the left shifts with the parameters $\epsilon^{ia}$ and
$\hat{\epsilon}^{ia}$, as well as the induced stability subgroup infinitesimal transformations,
can be found from the general formula
\be
    g^{-1}\left(\epsilon_{ia}Q^{ia}+\hat{\epsilon}_{ia}S^{ia}\right)g =
    g^{-1}\delta g + \omega_{ij}\,L^{ij} + \omega_{ab}\,R^{ab} +\omega_{(C)}\,C. \label{TranGen}
\ee
The explicit calculations yield the following transformations:
\bea
    && \delta \theta^{ia} =\epsilon^{ia} + 2im\,\theta^{ib}\theta^{ja}\hat{\epsilon}_{jb}\,, \qquad
    \delta  \hat{\theta}^{ia} =
\hat{\epsilon}^{ia} - 2im \left[\hat{\theta}^{j(b}\theta^{a)}_j\hat{\epsilon}^i_b +  \hat{\theta}^{(j}_b\theta^{i)b}\hat{\epsilon}^a_j\right],\nn
    && \delta  t =  -\,i \hat{\theta}^{ia} \hat{\epsilon}_{ia} - i \theta^{ia} \epsilon_{ia}
    + \frac{2m}{3}\,\theta^{ib}\theta^{ja}\theta_{jb}\hat{\epsilon}_{ia}\,.\label{tr}
\eea
The induced elements in \eqref{TranGen} are
\bea
    \omega_{ij} = -\,2im\, \theta^a_{(i}\,\hat{\epsilon}_{j)a}\,, \qquad
    \omega_{ab} = -\,2im\, \theta_{i(a}\,\hat{\epsilon}^{i}_{b)}\,,\qquad
    \omega_{(C)} = 2\,\theta^{ia}\hat{\epsilon}_{ia}\, .\label{trh}
\eea
It is straightforward to check that the coset-space Cartan forms undergo $SU(2)_{\rm L}\times SU(2)_{\rm R}$ induced transformations under the
coordinate transformations \eqref{tr}:
\bea
    \delta \left(\Delta\theta^{ia}\right) = \omega^i_{j}\,\Delta\theta^{ja}
+\omega^a_{b}\,\Delta\theta^{ib},\qquad
    \delta \big(\Delta\hat{\theta}^{ia}\big) = \omega^i_{j}\,\Delta\hat{\theta}^{ja}
+\omega^a_{b}\,\Delta\hat{\theta}^{ib}.
\eea

Superfields are assumed to transform according to the general law
\bea
    \delta  \Phi^A = \left(\omega_{ij}\,\tilde{L}^{ij}+\omega_{ab}\,\tilde{R}^{ab} +\omega_{(C)}\,\tilde{C}\right)^A_B\Phi^B\,,\label{SFtr}
\eea
where an external index $A$ of the superfield $\Phi^A$ specifies the $SU(2)_{\rm L}\times SU(2)_{\rm R}$ matrix representation
by which this superfield is transformed (and that of $\tilde{C}$).

The $SU(2|2)$ invariant ${\cal N}{=}\,8$, $d{=}1$ superspace integration
measure is given by
\bea
    d\zeta = dt\, d^4\theta\, d^4\hat{\theta}\,,\qquad  \quad \delta  \left(d\zeta\right) = 0\,.
\eea

\subsection{Chiral $SU(2|2)$ superspace}
We introduce the complex coordinates
\bea
    \zeta_{\rm c}= \left\lbrace t,\vartheta^{ia},\bar{\vartheta}^{jb}\right\rbrace ,\qquad
    \overline{(\vartheta_{ia})}=\bar{\vartheta}^{ia}, \label{complexbasis}
\eea
which are related to those defined in \eqref{sspace} as
\bea
    \vartheta^{ia} = \frac{1}{\sqrt{2}}\left(\theta^{ia}+i\hat{\theta}^{ia}
    - \frac{2im}{3}\,\hat{\theta}^{ib}\hat{\theta}^{ja}\hat{\theta}_{jb}
    \right),\qquad
    \bar{\vartheta}^{ia} = \frac{1}{\sqrt{2}}\left(\theta^{ia}-i\hat{\theta}^{ia} - \frac{2im}{3}\,\hat{\theta}^{ib}\hat{\theta}^{ja}\hat{\theta}_{jb}\right).
\eea
It will be also convenient to pass to the new infinitesimal parameters
\bea
    \eta^{ia} = \frac{1}{\sqrt{2}}\left(\epsilon^{ia}+i\hat{\epsilon}^{ia}\right),\qquad
    \bar{\eta}^{ia} = \frac{1}{\sqrt{2}}\left(\epsilon^{ia}-i\hat{\epsilon}^{ia}\right),\label{eta}
\eea
in terms of which the transformation properties of the superspace coordinates in \p{complexbasis} are as follows
\bea
    &&\delta\vartheta^{ia} = \eta^{ia} + 2m\,\vartheta^{ib}\vartheta^{ja}\left(\eta_{jb}-\bar{\eta}_{jb}\right),\qquad
    \delta\bar{\vartheta}^{ia} = \bar{\eta}^{ia}+2m\,\bar{\vartheta}^{ib}\bar{\vartheta}^{ja}\left(\eta_{jb}-\bar{\eta}_{jb}\right),\nn
    &&\delta t = -\,i\vartheta^{ia}\bar{\eta}_{ia} - i\bar{\vartheta}^{ia} \eta_{ia}
    + \frac{2m}{3}\,\theta^{ib}\theta^{ja}\theta_{jb}\hat{\epsilon}_{ia} + \frac{2m}{3}\,\hat{\theta}^{ib}\hat{\theta}^{ja}\hat{\theta}_{jb}\epsilon_{ia}\,.\label{c_tr}
\eea
The  measure of integration over \p{complexbasis} can be checked to be invariant under these transformations:
\bea
    d\zeta_c = dt\, d^4\vartheta\, d^4\bar{\vartheta},\qquad \delta \left(d\zeta_c\right) = 0\,.
\eea
Specializing to the $\eta^{i2}$-transformations in \eqref{c_tr} yields the odd transformations corresponding to the $SU(2|1)$ subgroup \eqref{leftSU2}.

The covariant derivatives \eqref{CovD} take the following form in the basis \p{complexbasis}:
\bea
    {\cal D}^{ia} &=& \partial^{ia} - i\bar{\vartheta}^{ia}\partial_{t} + i\left(\vartheta^{ia}
    - \bar{\vartheta}^{ia}\right)\tilde{C} - m\left(\vartheta^{ib}-\bar{\vartheta}^{ib}\right) \left(\vartheta^{ja}-\bar{\vartheta}^{ja}\right)\partial_{jb}\nn
    &&-\,m\left[\left(\vartheta^{a}_{j} - \bar{\vartheta}^{a}_{j}\right)\tilde{L}^{ij} - \left(\vartheta^{i}_{b} - \bar{\vartheta}^{i}_{b}\right)\tilde{R}^{ab}\right],\nn
    \bar{\cal D}^{ia} &=& \bar{\partial}^{ia} - i\vartheta^{ia}\partial_{t} + i\left(\vartheta^{ia}
    - \bar{\vartheta}^{ia}\right)\tilde{C}+m\left(\vartheta^{ib}-\bar{\vartheta}^{ib}\right) \left(\vartheta^{ja}-\bar{\vartheta}^{ja}\right)\bar{\partial}_{jb} \nn
    &&- \,m\left[\left(\vartheta^{a}_{j} - \bar{\vartheta}^{a}_{j}\right)\tilde{L}^{ij}
    - \left(\vartheta^{i}_{b} - \bar{\vartheta}^{i}_{b}\right)\tilde{R}^{ab}\right].\label{covD_complex}
\eea
Hereafter, we use the notation:
\bea
    \partial^{ia}=\partial/\partial \vartheta_{ia}\,,\qquad
    \bar{\partial}^{ia}=\partial/\partial \bar{\vartheta}_{ia}\,.
\eea

Now it is easy to show the existence of a left chiral subspace parametrized by the coordinates
\bea
    \zeta_L = \left\lbrace t_L,\vartheta^{ia}\right\rbrace,  \label{left}
\eea
where
\bea
    t_{L} = t - i\vartheta^{ia}\bar{\vartheta}_{ia} - im\left[\frac{1}{3}\, \bar{\vartheta}^{ib}\bar{\vartheta}^{ja}\bar{\vartheta}_{ia}
    - \frac{1}{2}\, \bar{\vartheta}^{ib}\vartheta^{ja}\bar{\vartheta}_{ia} - \frac{1}{2}\, \vartheta^{ib}\bar{\vartheta}^{ja}\bar{\vartheta}_{ia}
    +  \vartheta^{ib}\vartheta^{ja}\bar{\vartheta}_{ia}\right]\vartheta_{jb}\,.
\eea
Indeed, the set \p{left} is closed under the $SU(2|2)$ transformations
\bea
    \delta\vartheta^{ia} = \eta^{ia} + 2m\,\vartheta^{ib}\vartheta^{ja}\left(\eta_{jb}-\bar{\eta}_{jb}\right),\qquad
    \delta t_{L} = -\,2i\vartheta^{ia}\bar{\eta}_{ia} + \frac{4im}{3}\,\vartheta^{ib}\vartheta^{ja}\vartheta_{jb}\bar{\eta}_{ia}\,.
\eea
Actually, the set \p{left} can be identified with the following complex coset superspace of $SU(2|2)$:
\bea
    \frac{\left\lbrace\Pi^{ia},\bar{\Pi}^{jb},L^{ij},R^{ab},C,H\right\rbrace}{\left\lbrace\bar{\Pi}^{jb}, L^{ij},R^{ab},C\right\rbrace}\,.\label{coset_ch}
\eea

The invariant measure of integration over \p{left}, $d\zeta_{L}$, is defined by
\bea
    d\zeta_{L} = dt_{L}\,d^4\vartheta ,\qquad \delta \left(d\zeta_{L}\right) = 0\,.
\eea
In the coordinates \eqref{left}, the covariant derivatives \eqref{covD_complex} are written as
\bea
    {\cal D}^{ia} &=& \partial^{ia} - 2i\left[\bar{\vartheta}^{ia}+m\left(\vartheta^{ic}\vartheta^{ka}\bar{\vartheta}_{kc}
    -\frac{1}{3}\,\bar{\vartheta}^{ic}\bar{\vartheta}^{ka}\bar{\vartheta}_{kc}\right)\right]\partial_{t_L}
      - m\left(\vartheta^{ib}-\bar{\vartheta}^{ib}\right) \left(\vartheta^{ja}-\bar{\vartheta}^{ja}\right)\partial_{jb}\nn
    &&+\, i\left(\vartheta^{ia} - \bar{\vartheta}^{ia}\right)\tilde{C}-m\left[\left(\vartheta^{a}_{j} - \bar{\vartheta}^{a}_{j}\right)\tilde{L}^{ij}
    - \left(\vartheta^{i}_{b} - \bar{\vartheta}^{i}_{b}\right)\tilde{R}^{ab}\right],\nn
    \bar{\cal D}^{ia} &=& \bar{\partial}^{ia} + i\left(\vartheta^{ia} - \bar{\vartheta}^{ia}\right)\tilde{C}
    +m\left(\vartheta^{ib}-\bar{\vartheta}^{ib}\right) \left(\vartheta^{ja}-\bar{\vartheta}^{ja}\right)\bar{\partial}_{jb} \nn
    &&- \,m\left[\left(\vartheta^{a}_{j} - \bar{\vartheta}^{a}_{j}\right)\tilde{L}^{ij} - \left(\vartheta^{i}_{b}
    - \bar{\vartheta}^{i}_{b}\right)\tilde{R}^{ab}\right].
\eea
{}From the structure of the covariant derivative $ \bar{\cal D}^{ia}$ we observe that the general covariantly chiral $SU(2|2)$ superfield
$\Phi^A$,
\bea
    \bar{\cal D}^{ia}\Phi^A =0\,, \label{chiral}
\eea
can be made explicitly chiral after the appropriate $\vartheta, \bar\vartheta$-dependent $SU(2)_{\rm L}$ and $SU(2)_{\rm R}$ rotation of $\Phi^A$ with respect to
the external indices. For instance, if $\Phi^A$ has the $SU(2)_{\rm L}\times SU(2)_{\rm R}$ matrix assignment $\left({\bf 1/2, 1/2}\right)$, this additional
redefinition is given by
\bea
&& \Phi^{ia}\left(t,\vartheta_{ia},\bar{\vartheta}^{jb}\right) =
    \left[e^{-\frac{m}{2}\left(\vartheta_{jb}-\bar{\vartheta}_{jb}\right) \left(\vartheta^{b}_{k}-\bar{\vartheta}^{b}_{k}\right)\tilde{L}^{jk}}\right]^i_{\;l}
    \left[e^{\frac{m}{2}\left(\vartheta_{jb}-\bar{\vartheta}_{jb}\right) \left(\vartheta^{j}_{c}
    -\bar{\vartheta}^{j}_{c}\right)\tilde{R}^{bc}}\right]^a_{\;d}\,\Phi^{ld}_L\left(t_L,\vartheta_{ia}\right), \label{RedefChir} \\
&& \left(\tilde{L}^{jk}\right)^i_{\;l} = \frac{1}{2}\left(\varepsilon^{ij}\delta^{k}_{l} +\varepsilon^{ik}\delta^{j}_{l} \right),
\qquad \left(\tilde{R}^{bc}\right)^a_{\;d} = \frac{1}{2}\left(\varepsilon^{ab}\delta^{c}_{d} + \varepsilon^{ac}\delta^{b}_{d}\right).
\eea
On the other hand, it is not possible to eliminate $\tilde{C}$ from $\bar{\cal D}^{ia}$ in a similar way.
In fact, $\tilde{C}$ should always be vanishing on chiral superfields, as  follows from the anticommutation
relations \p{DD},  which are just the integrability conditions for the chirality constraint \eqref{chiral} and its anti-chirality counterpart:
\bea
    \bar{\cal D}^{ia}\Phi^A =0 \quad \Rightarrow \quad  \tilde{C}\Phi^A = 0\,.\label{chiral1}
\eea

\subsection{Harmonic superspace}
We perform a harmonization of the $SU(2)_{\rm L}$ indices and define the analytic subspace
\bea
    \zeta_{(A)}=\left\lbrace t_{(A)}, \theta^{+}_{a}, \hat{\theta}^{+}_{a},w^{\pm}_{i}\right\rbrace, \label{HSS}
\eea
where
\bea
    &&\theta^{+a}=\theta^{ia} w^{+}_{i},\qquad
    \hat{\theta}^{+a}=\hat{\theta}^{ia}w^{+}_{i} + im\,\hat{\theta}^{i}_{b}\hat{\theta}^{jb}\theta^{ka}w^{+}_{i}w^{+}_{j}w^{-}_{k},\nn
    &&t_{(A)}= t-i\theta^{ia}\theta^{k}_a w^{+}_{i}w^{-}_{k}-i\hat{\theta}^{ia}\hat{\theta}^{k}_a w^{+}_{i}w^{-}_{k}.
\eea
It is closed under the following $SU(2|2)$ transformations
\bea
    &&\delta \theta^{+}_a= \epsilon^{i}_aw^{+}_{i} - im\,\theta^{+}_b\theta^{+ b}\hat{\epsilon}^{i}_{a}w^{-}_{i}, \qquad
    \delta \hat{\theta}^{+}_a= \hat{\epsilon}^{i}_a w^{+}_{i} - i m\left(2\,\hat{\theta}^{+}_b\theta^{+b}\hat{\epsilon}^{i}_{a}
    - \hat{\theta}^{+}_b\hat{\theta}^{+b}\epsilon^{i}_{a}\right)w^{-}_{i},\nn
    &&\delta w^+_{i}= 2im\,\theta^{+}_a \hat{\epsilon}^{ja}w^{+}_{j}w^{-}_{i},\qquad \delta w^-_{i}=0\,,\qquad
    \delta t_{(A)} = 2i\left(\theta^{+}_a \epsilon^{ia} + \hat{\theta}^{+}_a\hat{\epsilon}^{ia}\right)w^{-}_{i}.
\eea
Note that the transformation properties of the harmonic variables $w^{\pm}_i$, as well as the precise relation between the ``central basis''
coordinates $(t, \theta^{ia}, \hat\theta^{ia})$ and the ``analytic basis'' coordinates  $(t_{(A)}, \theta^{+}_a, \hat\theta^{+}_a)\,$,
are uniquely fixed just by requiring \p{HSS} to be closed under the $SU(2|2)$ transformations.

For further calculations, it will be convenient to pass to another set of harmonic variables in the harmonic superspace \p{HSS},
\be
u^{+ i} = w^{+ i} -im\,\theta^+_a\hat\theta^{+ a} w^{-i}, \quad u^{-i} = w^{-i}. \lb{Newharm}
\ee
With this choice, the realization of the fermionic $SU(2|2)$ transformations in the analytic subspace is as follows,
\bea
&& \delta  u^{+ i} = \Lambda^{++} u^{-i}\,,\qquad \delta u^{-i} =0\,,\nn
&& \delta \theta^{+}_a = \epsilon^{+}_a + im\left(\hat{\theta}^{+}_b\theta^{+b}\epsilon^{-}_{a}-\theta^{+}_b\theta^{+ b}\hat{\epsilon}^{-}_{a}\right), \nn
&& \delta \hat{\theta}^{+}_a = \hat{\epsilon}^{+}_a + im\left(\hat{\theta}^{+}_b\hat{\theta}^{+b}\epsilon^{-}_{a}
-\hat{\theta}^{+}_b\theta^{+b}\hat{\epsilon}^{-}_{a}\right),\nn
&& \delta t_{(A)} = 2i\left(\theta^{+}_a \epsilon^{-a} + \hat{\theta}^{+}_a\hat{\epsilon}^{-a}\right),\lb{NewTrans}
\eea
where
\bea
    &&\Lambda^{++} = im\left(\theta^{+}_a \hat{\epsilon}^{+a}-\hat{\theta}^{+}_a\epsilon^{+a}\right)
    - m^2\left(\theta^{+}_a \hat{\epsilon}^{-a}+\hat{\theta}^{+}_a\epsilon^{-a}\right) \hat{\theta}^{+}_b\theta^{+b},\nn
    &&\epsilon^{\pm a}=\epsilon^{ia} u^{\pm}_{i},\qquad
    \hat{\epsilon}^{\pm a}=\hat{\epsilon}^{ia}u^{\pm}_{i}.
\eea
The $SU(2|2)$ covariant harmonic derivative ${\cal D}^{++}$ preserving analyticity is uniquely defined by requiring it to transform as
\be
    \delta {\cal D}^{++} = -\,\Lambda^{++}D^0, \qquad D^0 = u^{+i}\frac{\partial}{\partial u^{+ i}} - u^{-i}\frac{\partial}{\partial u^{- i}}
    +  \theta^{+a}\frac{\partial}{\partial \theta^{+ a}} +  \hat{\theta}^{+a}\frac{\partial}{\partial \hat{\theta}^{+ a}}\,.\label{D++Transf}
\ee
It reads
\be
    {\cal D}^{++} = \partial^{++} + i\left(\theta^+_a\theta^{+ a} + \hat{\theta}^+_a\hat{\theta}^{+ a}\right)\partial_A
+ im\,\hat{\theta}^{+}_b\theta^{+b}\left({\theta}^{+ a}\partial_{+ a} - \hat{\theta}^{+ a}\hat{\partial}_{+ a}\right)
    +\frac{m^2}{2}\left(\theta^{+}\right)^4\partial^{--},\label{calD}
\ee
where
$$
\partial^{\pm\pm} = u^{\pm i}\frac{\partial}{\partial u^{\mp i}}\,, \quad \partial_A  = \frac{\partial}{\partial t_{(A)}}\,, \quad
\partial_{+ a} = \frac{\partial}{\partial \theta^{+ a}}\,, \quad \hat{\partial}_{+ a} = \frac{\partial}{\partial \hat{\theta}^{+ a}}\,,\qquad\left(\theta^{+}\right)^4 :=
\left(\theta^+_b\theta^{+ b}\right)\left(\hat{\theta}^+_a\hat{\theta}^{+ a}\right).
$$
This harmonic derivative reveals some unusual properties to be used below:
\bea
    {\cal D}^{++} u^{-i} = u^{+i},\qquad {\cal D}^{++} u^{+ i} = \frac{m^2}{2}\left(\theta^{+}\right)^4\,u^{-i},
    \qquad {\cal D}^{++} \Lambda^{++} = \frac{m^2}{2}\,\delta \left(\theta^{+}\right)^4.\lb{DLambda++}
\eea

The analytic subspace integration measure
\be
d\zeta_{(A)}^{(-4)} := dt_{(A)}\, du\, d^2\theta^{+}\,d^2\hat{\theta}^{+}  \label{HSSmeasure}
\ee
transforms as
\be
    \delta\left(d\zeta_{(A)}^{(-4)}\right) =  2\,d\zeta_{(A)}^{(-4)}\,\Lambda\,, \lb{TranMeas}
\ee
with
\be
    \Lambda := im\left(\hat{\theta}^{+}_a\epsilon^{-a}-\theta^{+}_a\hat{\epsilon}^{-a}\right). \label{Lambda}
\ee
It is easy to check that
\be
\Lambda^{++} = -\, {\cal D}^{++} \Lambda\,.\label{DefLambdaLambda}
\ee

\subsection{Biharmonic superspace}
One can extend the superspace \eqref{sspace} by biharmonic coordinates $w^{(\pm 1)}_i$ and $v^{(\pm 1)}_a$
associated with the subgroups $SU(2)_{\rm L}$ and $SU(2)_{\rm R}$\,, respectively. No such an option exists in the
$SU(2|1)$ case because of presence of only one $SU(2)$ subgroup in $SU(2|1)$. Like in the previous case, the transformation
laws of the double set of harmonics and the precise proper change of the superspace coordinates can be found from the requirement
of the existence of the invariant analytic subspace $\zeta_{(B)}$ in the full biharmonic superspace
\bea
    \zeta_{(B)} = \left\lbrace t_{(B)}, \theta^{(1,0)}_a,\hat{\theta}^{(0,1)}_i, w^{(\pm 1)}_{i},v^{(\pm 1)}_{a} \right\rbrace. \label{zetaB}
\eea

The relation between the coordinates \eqref{zetaB} and the original coordinates  \eqref{sspace} is given by the following substitutions,
\bea
    &&\theta^{(1,0)}_{a}=\theta^{i}_{a} w^{(1)}_{i},\qquad
    \hat{\theta}^{(0,1)}_{i}=\left[\hat{\theta}^{a}_{i} - 2im \,\hat{\theta}^{b}_{i}\hat{\theta}^{ja}\theta_{jb}
    +4m^2\hat{\theta}^{b}_{i}\hat{\theta}^{ja}\hat{\theta}^{kc}\theta_{kb}\theta_{jb}\right]v^{(1)}_{a},\nn
    &&t_{(B)}= t-i\theta^{ia}\theta^{j}_a w^{(1)}_{i}w^{(-1)}_{j}-i\hat{\theta}^{ia}\hat{\theta}^{b}_i v^{(1)}_{a}v^{(-1)}_{b} + 2m\,\hat{\theta}^{ia}\hat{\theta}^{b}_i\hat{\theta}^{kc}\theta^{d}_{k} v^{(-1)}_{a}v^{(-1)}_{b}v^{(1)}_{c}v^{(1)}_{d}.
\eea
Now it is straightforward to explicitly find the relevant coordinate $SU(2|2)$ transformations leaving closed the analytic coordinate set \eqref{zetaB}
\bea
    &&\delta \theta^{(1,0)}_a = \epsilon^{(1,0)}_a - im\,\theta^{(1,0)}_b\theta^{(1,0)\,b}\hat{\epsilon}^{(-1,0)}_a , \qquad
    \delta \hat{\theta}^{(0,1)}_i= \hat{\epsilon}^{(0,1)}_i - im\,\hat{\theta}^{(0,1)}_j\hat{\theta}^{(0,1)\,j}\epsilon^{(0,-1)}_i,\nn
    &&\delta w^{(1)}_{i} = \Lambda^{(2,0)}\,w^{(-1)}_{i},\qquad \delta w^{(-1)}_{i}=0\,,\nn
    &&\delta v^{(1)}_{a} = \Lambda^{(0,2)}\,v^{(-1)}_{a},\qquad \delta v^{(-1)}_{a}=0\,,\nn
    &&\delta t_{(B)} = 2i\theta^{(1,0)}_a\epsilon^{(-1,0)\,a} + 2i\hat{\theta}^{(0,1)}_i \hat{\epsilon}^{(0,-1)\,i}.\label{tr_BHSS}
\eea
Here
\bea
    &&\epsilon^{(\pm 1,0)\,a} = \epsilon^{ia}w^{(1)}_{i},\quad
    \hat{\epsilon}^{(\pm 1,0)\,a} = \hat{\epsilon}^{ia}w^{(1)}_{i}, \qquad
    \epsilon^{(0,\pm 1)\,i} = \epsilon^{ia}v^{(\pm 1)}_{a},\quad
    \hat{\epsilon}^{(0,\pm 1)\,i} = \hat{\epsilon}^{ia}v^{(\pm 1)}_{a}, \label{epsilonbih} \\
    &&\Lambda^{(2,0)} = 2im\,\theta^{(1,0)}_a \hat{\epsilon}^{(1,0)\,a},\quad
\Lambda^{(0,2)} = 2im\,\hat{\theta}^{(0,1)}_i \epsilon^{(0,1)\,i}. \label{lambda2002}
\eea
The analytic subspace has an invariant integration measure,
\be
    d\zeta_{(B)}^{(-2,-2)} := dt_{(B)}\,dw\,dv\,d^2\theta^{(1,0)}\,d^2\hat\theta^{(0,1)},\qquad\delta\left(d\zeta_{(B)}^{(-2,-2)}\right) = 0\,.
    \label{measureB}
\ee

Now we can define the covariant harmonic derivatives
\bea
    &&D^{(2,0)}=\partial^{(2,0)} + i\theta^{(1,0)}_a\theta^{(1,0)\,a}\,\partial_{(B)} \,, \qquad
    D^{0}_w=\partial^{0}_w + \theta^{(1,0)\,a}\,\frac{\partial}{\partial\theta^{(1,0)\,a}}\,,\nn
    &&D^{(0,2)}=\partial^{(0,2)} + i\hat{\theta}^{(0,1)}_i\hat{\theta}^{(0,1)\,i}\,\partial_{(B)}\,,\qquad
    D^{0}_v=\partial^{0}_v + \hat{\theta}^{(0,1)\,i}\,\frac{\partial}{\partial\hat{\theta}^{(0,1)\,i}}\,,\label{D20}
\eea
where
\bea
    &&\partial^{(\pm 2,0)} = w^{(\pm 1)}_i\frac{\partial}{\partial w^{(\mp 1)}_i}\,, \qquad
    \partial^{0}_w = w^{(1)}_i\frac{\partial}{\partial w^{(1)}_i} - w^{(-1)}_i\frac{\partial}{\partial w^{(-1)}_i}\,,\qquad \partial_{(B)} = \partial / \partial t_{(B)}\,,\nn
    &&\partial^{(0,\pm 2)} = v^{(\pm 1)}_a\frac{\partial}{\partial v^{(\mp 1)}_a}\,,\qquad
    \partial^{0}_v = v^{(1)}_a\frac{\partial}{\partial v^{(1)}_a} - v^{(-1)}_a\frac{\partial}{\partial v^{(-1)}_a}\,.
\eea
They possess the standard transformation laws
\bea
    \delta D^{(2,0)}=-\,\Lambda^{(2,0)}\,D^{0}_w\,,\quad \delta D^0_w = 0\,, \qquad
    \delta D^{(0,2)}=-\,\Lambda^{(0,2)}\,D^{0}_v\,,\quad \delta D^0_v= 0\,. \label{transf2002}
\eea
One can check that
\bea
&& \Lambda^{(2,0)} = D^{(2,0)}\Lambda^{(0,0)},\quad
\Lambda^{(0,2)} = D^{(0,2)}\Lambda^{(0,0)}, \label{D20lambda00} \\
&& \Lambda^{(0,0)}= 2im\left[\epsilon^{(0,-1)}_{i}\hat{\theta}^{(0,1)\,i} + \hat{\epsilon}^{(-1,0)}_{a}\theta^{(1,0)\,a}\right].\label{lambda00}
\eea

\setcounter{equation}{0}
\section{The multiplet ${\bf (3,8,5)}$}\label{385}

\subsection{Kinematics}
On one hand, the multiplet ${\bf (3,8,5)}$ can be described by a superfield $V^{ij}$
satisfying the constraints
\bea
    D^{(i}_a V^{jk)}=0\,,\qquad \nabla^{(i}_a V^{jk)}=0\,,\qquad \tilde{C}\,V^{ij}=0\,.\label{constrCentr}
\eea
According to \eqref{SFtr}, the ``passive'' transformation law of $V^{ij}$ is
\bea
    \delta V^{ij}= \omega^{i}_{k}V^{kj} + \omega^{j}_{k}V^{ik}.
\eea
On the other hand, one can define the harmonic superfield
\bea
    V^{++}\left(\zeta_{(A)}\right) =
    V^{ij}w^{+}_{i}w^{+}_{j}\left[1+2im\left(\hat{\theta}^{ka}w^{+}_{k}
    +im\,\hat{\theta}^{k}_{b}\hat{\theta}^{mb}\theta^{na}w^{+}_{k}w^{+}_{m}w^{-}_{n}\right)\theta^{l}_{a}w^{-}_{l}\right],\label{V++}
\eea
which lives on the analytic harmonic superspace \eqref{HSS}. The Grassmann analyticity conditions for $V^{++}$ amount just to the constraints \p{constrCentr} for $V^{ij}$
in the original central basis. After passing to the new harmonic variables \eqref{Newharm}, the transformation of $V^{++}$
can be written through the parameter $\Lambda$ defined in \eqref{Lambda} as
\be
    \delta V^{++} = -\,2\Lambda V^{++}.    \label{trV++}
\ee
The analytic superfield $V^{++}$ satisfies the harmonic condition
\be
    {\cal D}^{++} V^{++} = 0\,,\label{ConstrV++}
\ee
which can be proved using \p{V++} expressed in terms of the harmonic $u^\pm_i\,$, as well as the explicit expression for ${\cal D}^{++}$, eq. \p{calD}. In fact
at this step one can forget about the relation \p{V++} and deal with the real analytic harmonic superfield $V^{++}$ ($\widetilde{V^{++}} = V^{++}$)  subjected to \p{ConstrV++}.
The harmonic constraint implies the following component structure of $V^{++}$:
\bea
    V^{++} & =& v^{ij}u^+_iu^+_j + \theta^{+a}\xi^i_a u^+_i + \hat{\theta}^{+a}\hat\xi^i_a u^+_i + \left(\theta^+_a\theta^{+a} - \hat{\theta}^+_a\hat{\theta}^{+a}\right) A_0 - i\dot{v}^{ij}u^+_iu^-_j\left(\theta^+_a\theta^{+a} + \hat{\theta}^+_a\hat{\theta}^{+a}\right)\nn
    &&-\, 2\,\theta^{+}_{a}\hat{\theta}^{+a} C_{0}  + \theta^{+ (a}\hat{\theta}^{+ b)} C_{ab}- i\theta^{+ a}\hat{\theta}^+_b\hat{\theta}^{+b}\left(\dot\xi^i_a + \frac{m}{2}\,\hat\xi^i_a\right)u^-_i
     -i\hat\theta^{+a}\theta^+_b\theta^{+b}\left(\dot{\hat\xi}^i_a - \frac{m}{2}\,\xi^i_a\right)u^-_i\nn
    && - \left(\theta^+\right)^4\left(\ddot{v}^{ij} + \frac{m^2}{2}\,v^{ij}\right)u^-_iu^-_j -\mu\left(\theta^+_a\theta^{+a}
    + \hat{\theta}^+_a\hat{\theta}^{+a}\right),\label{compv++}
\eea
where the fields satisfy the reality conditions:
\bea
    &&\overline{\left(v_{ij}\right)}=v^{ij},\qquad v^{ij}=v^{ji},\qquad \overline{\left(\xi_{ia}\right)}=-\,\xi^{ia},\qquad
    \overline{\big(\hat{\xi}_{ia}\big)}=-\,\hat{\xi}^{ia},\nn
    &&\overline{\left(C_{ab}\right)}=-\,C_{ab}\,,\qquad C_{ab}=C_{ba}\,,\qquad\overline{\left(C_{0}\right)}=-\,C_{0}\,,\qquad
    \overline{\left(A_{0}\right)}=-\,A_{0}\,.
\eea
Thus, we are left with three physical bosonic fields $v^{ij}(t)$, eight fermionic fields $\xi^i_a(t)$, $\hat\xi^i_a(t)$ and
five bosonic auxiliary fields $A_0(t)$, $C_0(t)$, $C_{ab}(t)$, {\it i.e.}, just with the $({\bf 3, 8, 5})$ content.
A new constant $\mu\,, \; [\mu ] =1\,,$ came out in the course of solving \p{ConstrV++}. It survives in the flat limit $m=0$.

We also present here the transformation properties of the component fields,
\bea
    &&\delta v^{ij} = \epsilon^{(i}_a \xi^{j)a}  + \hat{\epsilon}^{(i}_a \hat{\xi}^{j)a}\,, \nn
    && \delta \xi^i_a = 2i\left(\epsilon_{ja}\dot{v}^{ij} - m\,\hat{\epsilon}_{ja}v^{ij}\right)
    + 2\,\epsilon^i_a \left(A_0 - \mu\right) -2\,\hat{\epsilon}^{i}_{a}C_{0}
- \hat{\epsilon}^{ib}C_{ab}\,, \nn
    && \delta \hat{\xi}^i_a = 2i\left( \hat{\epsilon}_{ja} \dot{v}^{ij} + m\,\epsilon_{ja} {v}^{ij}\right)
    - 2\,\hat{\epsilon}^i_a\left(A_0 + \mu\right)-2\,\epsilon^i_a C_0
+ \epsilon^{ib}C_{ba}\,, \nn
    && \delta A_0 = \frac{i}{2}\left(\hat{\epsilon}_{ia}\dot{\hat\xi}^{ia}-\epsilon_{ia}\dot{\xi}^{ia}\right), \nn
    && \delta C_{0} =\frac{i}{2} \left(\epsilon_{ia}\dot{\hat{\xi}}^{ia} + \hat{\epsilon}_{ia}\dot{\xi}^{ia}\right),\nn
    && \delta C_{ab} = 2i \left[\hat{\epsilon}_{i(b}\dot{\xi}_{a)}^{i}-\epsilon_{i(a}\dot{\hat{\xi}}_{b)}^i \right]
+ 2im\left[\epsilon_{i(a}\xi_{b)}^i + \hat{\epsilon}_{i(a} \hat{\xi}_{b)}^i\right]. \label{TrnsfComp}
\eea

Note that the $SU(2|2)$ covariant constraint \p{ConstrV++} and the transformation law \p{trV++} can be generalized
to an arbitrary analytic superfield $q^{+n}$ of the harmonic
$U(1)$ charge $n$:
\be
    {\cal D}^{++} q^{(+n)} = 0\,,\quad
    \delta q^{(+n)}=-\,n\,\Lambda\,q^{(+n)}. \lb{DefCovSF}
\ee
This is similar to the analogous phenomenon observed in the flat ${\cal N}{=}\,4, d{=}1$ harmonic superspace \cite{IvLe}.
For even $n$ one can impose the reality condition on $q^{(+n)}$. The difference from the ${\cal N}{=}\,4, d{=}1$ case is that for $n=1$
the constraint in \p{DefCovSF} implies the equations of motion for the physical fields and is similar in this respect to the harmonic equation
of motion for the analytic hypermultiplet superfield in ${\cal N}{=}\,2, 4D$ case \cite{HSS0,HSS}. For $n=2\,,$ this constraint remains purely kinematic
and defines the $d{=}1$ analog of the ${\cal N}{=}\,2, d{=}4$ tensor multiplet, with the constant $\mu$ appearing as a solution of the $d{=}1$ reduction of the well-known
``notoph'' condition $\partial^\mu A_\mu = 0$ in $4D$. All these features are retained in the flat limit $m=0\,$.

\subsection{Invariant action}
Since for $n=2$ the $m=0$ version of \p{DefCovSF} defines an off-shell (${\bf 3, 8, 5}$) multiplet, it is expected
that the analytic real superfield $V^{++} :=q^{(+2)}$
describes the off-shell multiplet (${\bf 3, 8, 5}$) of $SU(2|2)$ supersymmetry as a deformation of the corresponding
flat off-shell supermultiplet.

Confronting the transformation law of $V^{++}$ \eqref{trV++} with the transformation of the analytic measure \p{TranMeas},
one concludes that it is impossible to construct any invariant Lagrangian
out of $V^{++}$ (even the free one), with harmonic $U(1)$ charge $+4$ (needed to cancel the negative charge $-4$ of the measure).
To evade this difficulty, we will proceed by analogy with the construction of the superconformal actions in \cite{GIO,IvLe}.

The procedure is as follows. We introduce an auxiliary constant triplet $c^{ij}$. Its harmonic projections
 $c^{+-} = c^{ij}u^+_i u^-_j$ and $c^{\pm\pm} =  c^{ij}u^{\pm}_i u^{\pm}_j$ satisfy the relation
\be
    c^{++}c^{--} - \left(c^{+-}\right)^2 = \frac{1}{2}\,c^{ij}c_{ij}
\ee
that follows from the completeness relation for the harmonics. Without loss of generality, we
choose $c^{ij}c_{ij} = 1\,$. Next, we define the ``shifted'' superfield $\hat{V}^{++}$ as
\be
    V^{++} = \hat{V}^{++} + \tilde{c}^{++}, \qquad \tilde{c}^{++} = c^{++} - \frac{m^2}{2}\left(\theta^+\right)^4 c^{--}.\lb{Vv}
\ee
The triplet $\tilde{c}^{++}$ satisfies the condition
\be
    {\cal D}^{++}\tilde{c}^{++} = 0\,,
 \ee
and so
\be
    {\cal D}^{++}\hat{V}^{++} = 0\,.\label{385constr}
\ee
The appearance of an additional term in $\tilde{c}^{++}$ is related to the properties \p{DLambda++} of ${\cal D}^{++}\,$.
Note also the useful relation
\be
    {\cal D}^{++} c^{+-} = c^{++} + \frac{m^2}{2}\left(\theta^+\right)^4 c^{--} = \tilde{c}^{++} + m^2\left(\theta^+\right)^4 c^{--}.
\ee
The component structure of the shifted analytic
superfield $\hat{V}^{++}$ related to $V^{++}$ by \p{Vv} is obtained from \p{compv++} just via the substitution $v^{ij} \rightarrow \hat{v}^{ij}\,,$
where $\hat{v}^{ij}= v^{ij} - c^{ij}$.

The newly defined quantities are transformed as
\bea
    &&\delta \tilde{c}^{++} = 4\Lambda^{++} c^{+-} - {\cal D}^{++}\left(\Lambda^{++} c^{--}\right),  \nn
    &&\delta \hat{V}^{++} = -\,2\Lambda \left(\hat{V}^{++} + \tilde{c}^{++}\right) - 4\Lambda^{++}c^{+-} + {\cal D}^{++}\left(\Lambda^{++} c^{--}\right),\lb{Tranvc}
\eea
where $\Lambda^{++} = -{\cal D}^{++}\Lambda$ (recall eq. \p{DefLambdaLambda}). Using these relations, one can construct invariant actions (see Appendix \ref{App. A.1})
with the superfield Lagrangian
\bea
    L^{(+4)} =\frac{2\,\big(\hat{V}^{++}\big)^2}{\left(1 + \sqrt{1 + 2\,c^{--}\hat{V}^{++}}\,\right)^2}
    -m^2 \left(\theta^+\right)^4\left(\frac{c^{--}\hat{V}^{++}}{\sqrt{1 + 2\,c^{--}\hat{V}^{++}}\,}
    + \frac{c^{--}\hat{V}^{++}}{1 + \sqrt{1 + 2\,c^{--}\hat{V}^{++}}\,}\right).\nn\lb{L(+4)}
\eea
To find the component form of the action
\bea
    S_{\bf (3,8,5)}=\int d\zeta^{(-4)}_{(A)} L^{(+4)}=\int dt\, {\cal L}_{\bf (3,8,5)}\,,\label{L385}
\eea
we use the normalization
\be
\int d\zeta_{(A)}^{(-4)} (\theta^+)^4 = \int dt\,du\,, \qquad d\zeta_{(A)}^{(-4)} =
\frac{1}{16}\,dt\,du\left(D^-_aD^{- a}\right)\left(\nabla^-_a\nabla^{- a}\right).
\ee
The main technical problem is to do the relevant harmonic integrals. This can be accomplished using the formulas listed in Appendix \ref{App. A.2}.
The component Lagrangian finally reads
\bea
    {\cal L}_{\bf (3,8,5)}&=& \frac{1}{2|v|}\,\bigg[\,\dot{v}_{ij}\dot{v}^{ij} +
\frac{i}{2}\left(\dot{\xi}^{ia}\xi_ {ia} + \dot{\hat{\xi}}^{ia}\hat{\xi}_ {ia}\right)
-\frac{i}{2|v|^2}\left(\xi^{(i}_a\xi^{j)a} + \hat{\xi}^{(i}_a\hat{\xi}^{j)a}\right)v_{ik}\dot{v}^k_j  \nn
    &&-\, \frac{v^{ij}}{2|v|^2}
\left(\xi^{a}_{i}\hat{\xi}^{b}_j\, C_{ab}-2\,\xi_{ia}\hat{\xi}^{a}_j\, C_{0}  + \xi_{ia}{\xi}^a_j\left(A_0 + \mu\right) - \hat{\xi}_{ia}\hat{\xi}^a_j
\left(A_0 - \mu\right)\right)-\frac{1}{4}\,C^{ab}C_{ab} \nn
    &&-\, 2 \left(A_0 + \mu\right) \left(A_0 - \mu\right)  - 2\left(C_0\right)^2  + \frac{3v_{(ij}v_{kl)}}{8|v|^4}\,\xi^a_{i}\xi_{ja}\,\hat{\xi}^b_{k}\hat{\xi}_{lb}
    + \frac{i}{2}\,m\,\hat{\xi}^{ia}\xi_{ia}-m^2v_{ij}v^{ij}\,\bigg]\nn
    &&-\,\frac{i\mu\,\dot{v}^{ij}\left(c^k_i v_{jk}+c^k_j v_{ik}\right)}
    {|v|\left(|v| + c_{ij} v^{ij}\right)}\,.\nn \label{KinComp1}
\eea
Here, $|v| := \sqrt{v_{ij}v^{ij}}$\,. The expression within the square brackets can basically be obtained by a dimensional reduction
$d{=}4 \rightarrow d{=}1$ from the $d{=}4$ Lagrangian of \cite{GIO}. The new terms are those $\sim \mu$ (they survive in the $m=0$ limit),
the fermionic ``mass'' mixed term $\sim m$ and the bosonic potential term $\sim m^2\,$. The last term $\sim\mu$ is a special
WZ term for $v^{ik}$ known as a Lorentz-force type coupling
to Dirac magnetic monopole \cite{IvLe,BIKL1}.

\subsection{Duality transformations}
In \cite{GIO}, duality transformations of the tensor multiplet was shown to lead to the free hypermultiplet action. Here, we define in the same way duality transformations
for the $d{=}1$ multiplet ${\bf (3,8,5)}$.

We can rewrite the action \eqref{L385} as
\bea
    S_{\bf (3,8,5)}=\frac{1}{2}\int d\zeta^{(-4)}_{(A)}\left(f^{++}\right)^2,\label{fPP}
\eea
where $f^{++}$ is an analytic superfield related to $V^{++}$ and $\hat{V}^{++}$ by
\bea
    &&f^{++} =\frac{2\hat{V}^{++}}{1 + \sqrt{1 + 2\,c^{--}\hat{V}^{++}}\,} - m^2 c^{--}\left(\theta^+\right)^4
    \left(1+\frac{1}{2\sqrt{1 + 2\,c^{--}\hat{V}^{++}}\,}\right)\quad\Rightarrow\nn
    &&V^{++} = f^{++}\left(1 + \frac{1}{2}\,c^{--}f^{++}\right) +c^{++}+ m^2 c^{--}\left(\theta^+\right)^4\left(1+c^{--}f^{++}\right),\nn
    &&\hat{V}^{++} = f^{++}\left(1 + \frac{1}{2}\,c^{--}f^{++}\right) + m^2 c^{--}\left(\theta^+\right)^4\left(\frac{3}{2}+c^{--}f^{++}\right).\label{f++}
\eea
In view of this one-to-one correspondence, the harmonic constraint \eqref{385constr} implies a nonlinear constraint on the superfield $f^{++}$.
The transformations of $f^{++}$ can be found from \p{f++}
\bea
    \delta f^{++}= \frac{2}{1 + c^{--}f^{++}}\left[-\,\Lambda f^{++}\left(1 + \frac{1}{2}\,c^{--}f^{++}\right)
    -\Lambda c^{++}-\Lambda^{++} c^{+-}\right]-2c^{--}{\cal D}^{++}\Lambda^{++}.\label{vf++}
\eea
Next, we add to the action \eqref{fPP} an additional term with the Lagrange multiplier $\omega$,
\bea
    S^{\rm dual}=\int d\zeta^{(-4)}_{(A)}\left[\frac{1}{2}\left(f^{++}\right)^2+\omega\,{\cal D}^{++}V^{++}\left(f^{++},u^{\pm}_i\right)\right],\label{DualOmV}
\eea
and thereby get rid of the condition \eqref{385constr}, ending up with two independent analytic superfields, $\omega$ and $f^{++}$.
The requirement of invariance of this action implies $\omega$ to transform as
\bea
    \delta \omega = -\,\frac{2\left(c^{+-}\Lambda + c^{--}\Lambda^{++}\right)}{1 + c^{--}f^{++}}\,.\label{omega}
\eea
Integrating by parts the last term in \p{DualOmV}, we obtain
\bea
    S^{\rm dual}=\int d\zeta^{(-4)}_{(A)}\left[\frac{1}{2}\left(f^{++}\right)^2 - V^{++}\left(f^{++},u^{\pm}_i\right){\cal D}^{++}\omega\right].\label{L385_1}
\eea
By analogy to \cite{GIO}, we can cast the Lagrangian \eqref{L385_1} in the form of the free action
\bea
    S^{\rm dual}=-\frac{1}{2}\int d\zeta^{(-4)}_{(A)}\, q^{+i}\,{\cal D}^{++}q^{+}_{i},\label{q+free}
\eea
where
\bea
    q^{+i} &:=& \left[f^{++}u^{-i} - 2 c^{ij}u^{+}_{j} + m^2 c^{--}u^{-i}\left(\theta^+\right)^4\right]\cos{\left(\omega/\sqrt{2}\right)}\nn
    &&-\,\sqrt{2}\left[c^{ij}f^{++}u^{-}_{j} + u^{+i} + m^2c^{ij}u^{-}_{j} c^{--}\left(\theta^+\right)^4\right]\sin{\left(\omega/\sqrt{2}\right)}. \label{q+}
\eea
{}From this relation, one can establish that
\bea
    V^{++}= \frac{1}{2}\,c_{ij}\,q^{+i}q^{+j}.
\eea
Taking into account \eqref{vf++} and \eqref{omega}, one can find the superfield transformation of the newly introduced analytic superfield $q^{+ i}$
\bea
    \delta q^{+i}=-\,\Lambda\,q^{+i}\,. \label{qtransf}
\eea
By making use of the transformation properties \p{D++Transf}, \p{TranMeas} and \p{qtransf}, it is easy to check the $SU(2|2)$ invariance of \eqref{q+free}.

We observe that the external doublet index $i$ of $q^{+ i}$ is inert with respect to the whole $SU(2|2)\,$, including the $SO(4)$ transformations.
So it a sort of Pauli-G\"ursey index and it is convenient to replace it by another letter, e.g. as
\bea
    q^{+i} \longrightarrow q^{+A}.
\eea
The action \eqref{q+free} respects an additional invariance under an extra $SU(2)_{\rm PG}$ rotating the doublet index $A$.

The superfield $q^{+A}$ has the following $\theta$-expansion:
\bea
    q^{+A}\left(\zeta_{(A)}\right) &=& x^{+A} + \theta^{+a}\lambda^A_a + \hat{\theta}^{+a}\hat{\lambda}^A_a + \left(\theta^+_a\theta^{+a} + \hat{\theta}^+_a\hat{\theta}^{+a}\right)B^{-A}+\left(\theta^+_a\theta^{+a} - \hat{\theta}^+_a\hat{\theta}^{+a}\right)C^{-A}\nn
    && +\,\theta^{+}_{a}\hat{\theta}^{+a}D^{-A}
    + \hat{\theta}^+_b\hat{\theta}^{+b}\theta^{+a}\psi^{(-2)A}_a + \theta^+_b\theta^{+b}\hat{\theta}^{+a}\hat{\psi}^{(-2)A}_a + \left(\theta^+\right)^4 A^{(-3)A} .
\eea
Here, all fields are defined on the extended bosonic space $\left(t_{(A)}, u^{\pm i}\right)$,
{\it i.e.}, their harmonic expansions produce infinite towers of fields \cite{HSS}.
Eliminating auxiliary fields by the relevant part of the equation of motion for \eqref{q+free},
\bea
    {\cal D}^{++}q^{+A}=0\,, \label{D++q+}
\eea
we obtain the on-shell superfield $q^{+A}$ containing a finite set of physical fields,
\bea
    && q^{+A} = x^{iA}u^+_i + \theta^{+a}\lambda^A_a + \hat{\theta}^{+a}\hat\lambda^A_a - i\left(\theta^+_a\theta^{+a}
    + \hat{\theta}^+_a\hat{\theta}^{+a}\right)\dot{x}^{iA}u^{-}_i ,\\
    &&\overline{\left(x_{iA}\right)}=x^{iA},\qquad \overline{\left(\lambda_{Aa}\right)}=-\,\lambda^{Aa},\qquad
    \overline{\big(\hat{\lambda}_{Aa}\big)}=-\,\hat{\lambda}^{Aa}.
\eea
The constraint \eqref{D++q+} puts the residual component fields on-shell:
\bea
    \ddot{x}^{iA} + \frac{m^2}{4}\,x^{iA}=0\,,\qquad
    \dot{\lambda}^{Aa}+\frac{m}{2}\,\hat{\lambda}^{Aa}=0\,,\qquad \dot{\hat{\lambda}}^{Aa}-\frac{m}{2}\,\lambda^{Aa} = 0\,.
\eea
They can be re-derived from the on-shell component Lagrangian
\be
    {\cal L}^{\rm dual}=\frac{1}{2}\,\dot{x}^{iA}\dot{x}_{iA} +\frac{i}{4}\left(\dot{\lambda}^{Aa}\lambda_{Aa}+\dot{\hat{\lambda}}^{Aa}\hat{\lambda}_{Aa}\right)
    +\frac{i}{4}\,m\,\hat{\lambda}^{Aa}\lambda_{Aa} - \frac{m^2}{8}\,x^{iA}x_{iA}\,, \label{OnSh}
\ee
which is invariant under the transformations
\bea
    \delta x^{iA} = \epsilon^{i}_a \lambda^{Aa}  + \hat{\epsilon}^{i}_a \hat{\lambda}^{Aa},\qquad
    \delta \lambda^{Aa} = 2i\epsilon^a_{i}\dot{x}^{iA}-im\,\hat{\epsilon}^a_{i}x^{iA},\qquad
    \delta \hat{\lambda}^{Aa} = 2i \hat{\epsilon}^a_{i} \dot{x}^{iA} +im\,\epsilon^a_{i}x^{iA}.
\eea
These have an on-shell $SU(2|2)$ closure. Note that the translational symmetry $x^{iA}\rightarrow x^{iA} + a^{iA}$ and its fermionic counterpart
which are present in the undeformed $m=0$ version of \p{OnSh} are broken by the oscillator terms.

\setcounter{equation}{0}
\section{The multiplet ${\bf (4,8,4)}$}\label{484}

\subsection{Kinematics}
The multiplet ${\bf (4,8,4)}$ can be described by the superfield $q^{ia}$, with $\overline{\left(q_{ia}\right)} = q^{ia}$.
The proper constraints are imposed as
\begin{equation}
    D^{(kb}q^{i)a}=0\,,\qquad \nabla^{k(b}q^{ia)}=0\,.\label{constr484}
\end{equation}
The $SU(2|2)$ covariance of these constraints requires that
\bea
    \tilde{C} q^{ia} = -\,im\,q^{ia}. \label{Cqia}
\eea
According to \eqref{SFtr}, the odd transformations of $q^i$ can be written as
\bea
    \delta q^{ia} = 2im\left(\hat{\epsilon}_{jb}\theta^{ib}q^{ja} + \hat{\epsilon}^{ja}\theta_{jb}q^{ib}\right).
\eea

Now one can define the analytic biharmonic superfield
\bea
    q^{(1,1)}\left(\zeta_{(B)}\right) = q^{ia}w^{(1)}_{i}v^{(1)}_{a}
    + 2im\,\hat{\theta}^{(0,1)}_{k}\theta^{k}_{b}q^{ib}w^{(1)}_{i}, \label{q11}
\eea
living on the analytic subspace \eqref{zetaB} and transforming as
\bea
    \delta q^{(1,1)} = \Lambda^{(0,0)}q^{(1,1)}, \label{tr_q}
\eea
where $\Lambda^{(0,0)}$ was defined in \p{lambda00}.  While the Grassmann constraints \p{constr484} are automatically satisfied for $q^{(1,1)}$ in the analytic basis,
the restricted harmonic dependence in \p{q11} amounts to the harmonic constraints
\bea
    D^{(2,0)}q^{(1,1)} = D^{(0,2)}q^{(1,1)}=0\,.\label{AC}
\eea
Taking into account the transformation laws of $D^{(2,0)}$ and $D^{(0,2)}$, eqs. \p{transf2002}, as well as the definitions \p{D20lambda00} and \p{lambda00},
it is easy to establish the $SU(2|2)$ covariance of \p{AC}.

The solution of \eqref{AC} is given by the undeformed superfield
\bea
    q^{(1,1)} &=& f^{ia} w^{(1)}_i v^{(1)}_a + \theta^{(1,0)\,a} v^{(1)}_b \chi^{\;\;b}_a + \hat{\theta}^{(0,1)\,i} w^{(1)}_j
    \hat{\chi}^{\;\;j}_i + \theta^{(1,0)\,a}\hat{\theta}^{(0,1)\,i}F_{ia}\nn
    && - \,i\left(\theta^{(1,0)}_b \theta^{(1,0)\,b} v^{(1)}_a w^{(-1)}_i +  \hat{\theta}^{(0,1)}_j\hat{\theta}^{(0,1)\,j} v^{(-1)}_a w^{(1)}_i\right)\dot{f}^{ia}\nn
    && - \,i\hat{\theta}^{(0,1)}_j\hat{\theta}^{(0,1)\,j}\theta^{(1,0)\,a} v^{(-1)}_b \dot{\chi}^{\;\;b}_a
    - i\theta^{(1,0)}_b\theta^{(1,0)\,b}\hat{\theta}^{(0,1)\,i} w^{(-1)}_j \dot{\hat{\chi}}^{\;\;j}_i \nn
    && - \,\hat{\theta}^{(0,1)}_j\hat{\theta}^{(0,1)\,j}\theta^{(1,0)}_b\theta^{(1,0)\,b} w^{(-1)}_i v^{(-1)}_a\ddot{f}^{\,ia}.\label{q11comp}
\eea
With taking into account \eqref{tr_q} and \eqref{tr_BHSS}, its components transformations are found to read
\bea
    &&\delta f^{ia} = -\,\epsilon^{ib}\chi^{\;\;a}_b - \hat{\epsilon}^{ja}\hat{\chi}^{\;\;i}_j \,,\qquad
    \delta F_{ia} =  2i\left(\epsilon_{ja}\dot{\hat{\chi}}^{\;\;j}_i-\hat{\epsilon}_{ib}\dot{\chi}^{\;\;b}_a\right)+
    2im\left(\hat{\epsilon}_{ja}\hat{\chi}^{\;\;j}_i-\epsilon_{ib}\chi^{\;\;b}_a\right),\nn
    &&\delta\chi^{\;\;b}_a = 2i\left(\epsilon_{ia}\dot{f}^{ib} - m\,\hat{\epsilon}_{ia}f^{ib}\right)-\hat{\epsilon}^{ib}F_{ia}\,,\qquad
    \delta\hat{\chi}^{\;\;j}_i = 2i\left(\hat{\epsilon}_{ia}\dot{f}^{ja} - m\,\epsilon_{ia}f^{ja}\right)+\epsilon^{ja}F_{ia}\,.\label{484tr}
\eea

Since the superfield $q^{(1,1)}$ in itself is not deformed (only its transformation properties prove to be deformed),
we can realize on it the supersymmetry $SU(2|2)$ in parallel with the standard flat ${\cal N}{=}\,8, d{=}1$ Poincar\'e supersymmetry, or
even with another $SU(2|2)$ involving the flipped-sign deformation parameter $-m$. The closure of all these symmetries including the original $SU(2|2)$
turns out to constitute an extended superalgebra introduced in \cite{BHSS}:
\bea
    &&\left\lbrace Q^0_{ia}, \tilde{S}_{jb}\right\rbrace = \varepsilon_{ij}\,J_{ab} - \varepsilon_{ab}\,I_{ij} + \frac{1}{2}\,\varepsilon_{ab}\,\varepsilon_{ij}\,Z,\nn
    &&\left\lbrace S^0_{ia}, \tilde{Q}_{jb}\right\rbrace = \varepsilon_{ab}\,J_{ij} - \varepsilon_{ij}\,I_{ab} + \frac{1}{2}\,\varepsilon_{ab}\,\varepsilon_{ij}\,Z,\nn
    &&\left\lbrace Q^0_{ia}, Q^0_{jb}\right\rbrace = 2\,\varepsilon_{ij}\,\varepsilon_{ab}\,H,\qquad \left\lbrace S^0_{ia}, S^0_{jb}\right\rbrace
    = 2\,\varepsilon_{ij}\,\varepsilon_{ab}\,H,\label{ext_a1}
\eea
\bea
    &&\left[I_{ij}, I_{kl}\right] = \varepsilon_{il}\,I_{kj} +\varepsilon_{jk}\,I_{il}\,, \qquad \left[I_{ab}, I_{cd}\right]
    = \varepsilon_{ad}\,I_{bc} +\varepsilon_{bc}\,I_{ad}\,,\nn
    &&\left[J_{ij}, J_{kl}\right] = \varepsilon_{il}\,J_{kj} +\varepsilon_{jk}\,J_{il}\,, \qquad \left[J_{ab}, J_{cd}\right]
    = \varepsilon_{ad}\,J_{bc} +\varepsilon_{bc}\,J_{ad}\,,
\eea
\bea
    &&\left[I_{ab}, S^0_{kc}\right] = \frac{1}{2}\left(\varepsilon_{ac}\,S^0_{kb} + \varepsilon_{bc}\,S^0_{ka}\right), \qquad
    \left[I_{ij}, Q^0_{ka}\right] = \frac{1}{2}\left(\varepsilon_{ik}\,Q^0_{ja} + \varepsilon_{jk}\,Q^0_{ia}\right),\nn
    &&\left[I_{ab}, \tilde{Q}_{kc}\right] = \frac{1}{2}\left(\varepsilon_{ac}\,\tilde{Q}_{kb} + \varepsilon_{bc}\,\tilde{Q}_{ka}\right),\qquad
    \left[I_{ij}, \tilde{S}_{ka}\right] = \frac{1}{2}\left(\varepsilon_{ik}\,\tilde{S}_{ja} + \varepsilon_{jk}\,\tilde{S}_{ia}\right),\nn
    &&\left[J_{ab}, Q^0_{kc}\right] = \frac{1}{2}\left(\varepsilon_{ac}\,Q^0_{kb} + \varepsilon_{bc}\,Q^0_{ka}\right),\qquad
    \left[J_{ij}, S^0_{ka}\right] = \frac{1}{2}\left(\varepsilon_{ik}\,S^0_{ja} + \varepsilon_{jk}\,S^0_{ia}\right),\nn
    &&\left[J_{ab}, \tilde{S}_{kc}\right] = \frac{1}{2}\left(\varepsilon_{ac}\,\tilde{S}_{kb} + \varepsilon_{bc}\,\tilde{S}_{ka}\right),\qquad
     \left[J_{ij}, \tilde{Q}_{ka}\right] = \frac{1}{2}\left(\varepsilon_{ik}\,\tilde{Q}_{ja} + \varepsilon_{jk}\,\tilde{Q}_{ia}\right).
    \label{ext_a2}
\eea
In fact, the superalgebra \eqref{ext_a1} -- \eqref{ext_a2} contains four $SU(2)$ subalgebras with the generators $I_{ab}$\,, $I_{ij}$\,, $J_{ab}$\,, $J_{ij}$.
These generators differently act on the indices of the component fields $\left\lbrace f^{ia},\chi^{\;\;a}_b ,\hat{\chi}^{\;\;i}_j, F_{ia}\right\rbrace$.
The generators $I_{ab}$\,, $I_{ij}$ rotate only the upper-case indices $i$ and $a$, while $J_{ab}$\,, $J_{ij}$ act only the lower-case ones (though denoted
by the same characters). Thus, the two types of $SU(2)$ indices of the component fields can actually be split into four types.

The $SU(2|2)$ generators of \eqref{algebra} can be identified with the following linear combinations of the generators
of the extended superalgebra \eqref{ext_a1} -- \eqref{ext_a2}:
\bea
    &&Q_{ia}=Q^0_{ia}-2im\,\tilde{Q}_{ia}\,,\qquad
    S_{ia}=S^0_{ia}-2im\,\tilde{S}_{ia}\,,\nn
    &&L_{ij}=-\left(J_{ij}+I_{ij}\right),\qquad R_{ab}=-\left(J_{ab}+I_{ab}\right).\label{new}
\eea
Hence, the superalgebra \eqref{algebra} can be viewed as a subalgebra of the extended superalgebra \eqref{ext_a1} -- \eqref{ext_a2},
with the central charge
\be
C = -\,imZ. \label{CZ}
\ee
The second $SU(2|2)$ supergroup is generated by the supercharges
\bea
    Q^0_{ia}+2im\,\tilde{Q}_{ia}\,,\qquad
    S^0_{ia}+2im\,\tilde{S}_{ia}\,.
\eea
The integration measure \eqref{measureB} is invariant under the transformations of both $SU(2|2)$ supergroups, with the parameters $m$ and $-m\,$,
{\it i.e.} it is also invariant under the transformations produced by all generators of \eqref{ext_a1} -- \eqref{ext_a2}.

The generators  appearing in \eqref{new}
are realized on the biharmonic superspace \eqref{zetaB} as
\bea
    Q^0_{ia} &=& w^{(1)}_i\frac{\partial}{\partial \theta^{(1,0)\,a}}
    - 2i w^{(-1)}_i \theta^{(1,0)}_a\partial_{(B)} \,, \quad S^0_{ia} = v^{(1)}_a\frac{\partial}{\partial \hat{\theta}^{(1,0)\,i}}
    - 2i v^{(-1)}_a \hat{\theta}^{(0,1)}_i\partial_{(B)}\,,\nn
    \nn
    \tilde{Q}_{ia}&=&\hat{\theta}^{(0,1)}_i v^{(1)}_a\partial^{(0,-2)}
    + \frac{1}{2}\,\hat{\theta}^{(0,1)}_j\hat{\theta}^{(0,1)\,j}\,v^{(-1)}_a\,\frac{\partial}{\partial \hat{\theta}^{(0,1)\,i}}-\hat{\theta}^{(0,1)}_i v^{(-1)}_a Z,\nn
    \tilde{S}_{ia} &=& \theta^{(1,0)}_a w^{(1)}_i\partial^{(-2,0)}
    + \frac{1}{2}\,\theta^{(1,0)}_b\theta^{(1,0)\,b}\,w^{(-1)}_i\frac{\partial}{\partial \theta^{(1,0)\,a}} - \theta^{(1,0)}_a w^{(-1)}_i Z,\nn
    I_{ij} &=& w^{(1)}_i w^{(1)}_j \partial^{(-2,0)}
    + \frac{1}{2}\left[w^{(1)}_{i} w^{(-1)}_{j} + w^{(1)}_{j} w^{(-1)}_{i}\right]\left(\theta^{(1,0)\,a}\frac{\partial}{\partial \theta^{(1,0)\,a}} - Z\right)\nn
    && +\, i w^{(-1)}_i w^{(-1)}_j\theta^{(1,0)}_a\theta^{(1,0)\,a}\,\partial_{(B)}\,,\nn
    I_{ab} &=& v^{(1)}_a v^{(1)}_b \partial^{(0,-2)} + \frac{1}{2}\left[v^{(1)}_{a} v^{(-1)}_{b} + v^{(1)}_{b} v^{(-1)}_{a}\right]
    \left(\hat{\theta}^{(0,1)\,k}\frac{\partial}{\partial \hat{\theta}^{(0,1)\,k}} - Z\right)\nn
    &&+\, i v^{(-1)}_a v^{(-1)}_b\hat{\theta}^{(0,1)}_i\hat{\theta}^{(0,1)\,i}\,\partial_{(B)} \,,\nn
    J_{ij} &=& -\,
    \frac{1}{2}\Bigg[\hat{\theta}^{(0,1)}_i\frac{\partial}{\partial \hat{\theta}^{(0,1)\,j}}+\hat{\theta}^{(0,1)}_j\frac{\partial}{\partial \hat{\theta}^{(0,1)\,i}}\Bigg],
    \,
    J_{ab} = -\,\frac{1}{2}\Bigg[\theta^{(1,0)}_a\frac{\partial}{\partial \theta^{(1,0)\,b}}+\theta^{(1,0)}_b\frac{\partial}{\partial \theta^{(1,0)\,a}}\Bigg].\label{ConCr}
\eea
While applying these operators to the superfield $q^{(1,1)}$, one is led to put $Z q^{ia} = q^{ia}$, in accord with \p{Cqia} and \p{CZ}. The algebra of the generators
\p{ConCr} can be extended by the generator
\be
K = i\,\theta^{(1,0)}_a \theta^{(1,0)a}\,\partial^{(-2,0)} + i\,\theta^{(0,1)}_i\theta^{(0,1)i}\,\partial^{(0,-2)} - t_{(B)} Z\,,\label{Kdefin}
\ee
which, together with $H = i\partial_{(B)}$ and $Z\,,$ form the Heisenberg algebra ${\bf h}(2)$
\be
[K, H] = iZ\,. \label{Heis}
\ee
The superalgebra \eqref{ext_a1} -- \eqref{ext_a2} with the generator $K$ being included can be treated as ${\cal N}{=}\,8$ extension of the algebra ${\bf h}(2)$ \cite{BHSS}.

\subsection{Invariant actions}
Let us define the new ``shifted'' superfield
\bea
    &&\hat{q}^{(1,1)}=q^{(1,1)}-c^{(1,1)},\qquad X = 2\,c^{(-1,-1)}\hat{q}^{(1,1)},\nn
    &&c^{(1,1)}=c^{ia}w^{(1)}_i v^{(1)}_a,\quad c^{(-1,-1)}=c^{ia}w^{(-1)}_i v^{(-1)}_a,\quad c^{(1,1)}c^{(-1,-1)}-c^{(-1,1)}c^{(1,-1)} = \frac{1}{2}\,,
\eea
where $c^{ia}$ is a constant satisfying $c^{ia}c_{ia}=1$. It is enough to consider the $\epsilon$-transformations
\bea
    &&\delta\,\hat{q}^{(1,1)} = \Lambda^{(0,0)}_{\epsilon}\hat{q}^{(1,1)} + \Lambda^{(0,0)}_{\epsilon}c^{(1,1)} - \Lambda^{(0,2)}_{\epsilon}c^{(1,-1)},\nn
    &&\delta\, X = \Lambda^{(0,0)}_{\epsilon}X + 2\,c^{(-1,-1)}c^{(1,1)}\Lambda^{(0,0)}_{\epsilon} - 2\,c^{(-1,-1)}c^{(1,-1)}\Lambda^{(0,2)}_{\epsilon}.
\eea
Such transformations are similar to the ``superconformal'' transformations \cite{BHSS}.
Then it follows that an $SU(2|2)$ invariant action can be constructed in the same way:
\bea
    S_{\bf (4,8,4)} = \int dt {\cal L}_{\bf (4,8,4)}= \int d\zeta_{(B)}^{(-2,-2)}\,\hat{q}^{(1,1)}\hat{q}^{(1,1)}
    \left[\frac{\ln{\left(1+X\right)}}{X^2}-\frac{1}{\left(1+X\right)X}\right].
    \label{484L}
\eea
Since the superfield $\hat{q}^{(1,1)}$ is not deformed, this action coincides with the one given in \cite{BHSS}
and so it is invariant under the full hidden supersymmetry with the algebra \eqref{ext_a1} -- \eqref{ext_a2} and the additional
transformations with the generator \p{Kdefin}. The central charge generator $Z$ acts as a dilatation generator
in the target space, $\delta_{Z}q^{(1,1)} = \omega q^{(1,1)}\,,$ where $\omega$ is a constant parameter. Note that \p{484L} is not invariant
under the standard dilatations which affect not only $q^{(1,1)}$, but also the time coordinate $t_{(B)}$.

Despite the transformations \eqref{484tr} are mass-deformed, the component Lagrangian of \eqref{484L} contains no terms with the parameter $m$.
In particular, the bosonic core of this Lagrangian is as follows
\bea
    {\cal L}_{\bf (4,8,4)}^{\rm bos} = \frac{1}{f^{2}}\left(\dot{f}_{ia}\dot{f}^{ia} - \frac{1}{4}\,F_{ia} F^{ia}\right),\qquad f^2=f_{ia}f^{ia}.\label{bosCore}
\eea
On the other hand, from the $SU(2|1)$ standpoint, the multiplet $({\bf 4, 8, 4})$ is a direct sum of two $SU(2|1)$ multiplets,
$({\bf 4, 8, 4}) =({\bf 4, 4, 0}) \oplus ({\bf 0, 4, 4})\,$, and it is known \cite{DHSS} that the Lagrangian of the multiplet $({\bf 4, 4, 0})$
in the general case explicitly involves the deformation parameter $m$. In particular, its bosonic core is
$$
\sim G (f)\Big( \dot{f}^{ia}\dot{f}_{ia} - \frac{m^2}{4}{f}^{ia}f_{ia}\Big).
$$
The only option for which the mass term becomes a constant and so fully decouples is just $G(f) = 1/f^2$ required by the $SU(2|2)$ invariance.\footnote{
Note that the singularity of this metric at $f^{ia} = 0$ can be avoided, e.g., by assuming that $f^{ia}$ starts with a constant
$c^{ia}$. The 4-dimensional manifold with such a metric is known as Hopf manifold (see e.g. \cite{Smilga1}) and it provides a simplest non-trivial example of the
so called HKT (``hyper-K\"ahler with torsion'') manifolds.}
It can be shown that for this special choice the parameter $m$ disappears also from all other terms in the $({\bf 4, 4, 0})$ Lagrangian.

Note that the Lagrangian \p{bosCore} is invariant under the $Z$ `dilatations'', $\delta_{Z} f^{ia} = \omega f^{ia}, \delta F^{ia} = \omega F^{ia}$ and,
up to a total derivative, under the transformations generated by the operator $K$ defined in \p{Kdefin}, $\delta_{K}f^{ia} = \omega'\,t f^{ia}\,, \;
\delta_{K}F^{ia} = \omega'\,t F^{ia}\,$.

\setcounter{equation}{0}
\section{The multiplet ${\bf (5,8,3)}$}\label{583}

\subsection{Kinematics}
In the standard flat ${\cal N}{=}\,8, d{=}1$ supersymmetry, chiral superfields with some extra constraints can be used to describe the supermultiplets ${\bf (2,8,6)}$
and ${\bf (5,8,3)}$ (see the table of ${\cal N}{=}\,8$ supermultiplets in \cite{BIKL2}).
As we have checked, the constraints defining the multiplet ${\bf (2,8,6)}$ do not admit a generalization to the $SU(2|2)$ supersymmetry, while
those defining the supermultiplet ${\bf (5,8,3)}$ can be $SU(2|2)$ covariantized. Thus, the supermultiplet ${\bf (5,8,3)}$ proves to be
the only $SU(2|2)$  multiplet for the description of which a chiral superfield can be utilized.

We consider the complex superfield $\Psi$ satisfying the standard chiral constraints
\bea
    \bar{\cal D}^{ia}\Psi = 0\,,\qquad \tilde{L}^{ij}\Psi = \tilde{R}^{ab}\Psi = \tilde{C}\,\Psi.\label{ch}
\eea
This superfield lives as unconstrained on the chiral subspace \eqref{left}. It means that the solution of \eqref{ch} is given by the general $\vartheta$-expansion
\bea
    &&\Psi\left(t_L,\vartheta^{ia}\right) = z + \sqrt{2}\,\vartheta^{ia}\psi_{ia} + \vartheta^{ia}\vartheta^{j}_{a}A_{ij} + \vartheta^{ia}\vartheta^{b}_{i}B_{ab} +
    \frac{2\sqrt{2}}{3}\,\vartheta^{ib}\vartheta^{ja}\vartheta_{jb}\,\pi_{ia}+\frac{1}{3}\,\vartheta^{ib}\vartheta^{ja}\vartheta_{jb}\vartheta_{ia}D,\nn
    &&B_{ab}=B_{ba}\,,\qquad A_{ij}=A_{ji}\,.\label{Phi}
\eea
The passive transformation law $\delta \Psi = 0$ implies the following component transformations:
\bea
    &&\delta z = -\,\sqrt{2}\,\eta^{ia}\psi_{ia}\,,\qquad
    \delta \psi_{ia} = -\,\sqrt{2}\left(\eta^j_a A_{ij} + \eta^b_i B_{ab} - i\bar{\eta}_{ia} \dot{z}\right),\nn
    &&\delta A_{ij} = -\,\sqrt{2}\,\eta^a_{(j}\left[\pi_{i)a}+m\,\psi_{i)a}\right] - \sqrt{2}\,\bar{\eta}^a_{(j}\left[i\dot{\psi}_{i)a} - m\,\psi_{i)a}\right],\nn
    &&\delta B_{ab} = -\,\sqrt{2}\,\eta_{i(b}\left[\pi^{i}_{a)}+m\,\psi^{i}_{a)}\right] + \sqrt{2}\,\bar{\eta}_{i(b}\left[i\dot{\psi}^{i}_{a)} + m\,\psi^{i}_{a)}\right],\nn
    &&\delta\pi_{ia} =\sqrt{2}\left(-\,i\bar{\eta}^j_a \dot{A}_{ij} + i\bar{\eta}^b_i \dot{B}_{ab} - \eta_{ia}D\right) +\sqrt{2}\,m\left[\left(\eta^j_a - \bar{\eta}^j_a\right) A_{ij} + \left(\eta^b_i - \bar{\eta}^b_i\right) B_{ab} - i\bar{\eta}_{ia}\dot{z}\right],\nn
    &&\delta D = \sqrt{2}\,i\bar{\eta}^{ia}\left(\dot{\pi}_{ia} + m\,\dot{\psi}_{ia}\right).\label{ch_tr}
\eea
Indeed, their Lie brackets are easily checked to form $SU(2|2)$ symmetry. The chiral superfield \eqref{Phi} contains $16$ bosonic and $16$ fermionic fields
and so is reducible. To single out the multiplet ${\bf (5,8,3)}$, we impose the extra $SU(2|2)$ covariant constraints
\bea
    &&\bar{\cal D}^{ia}\bar{\cal D}^{b}_{i}\,\bar{\Psi} + {\cal D}^{ia}{\cal D}^{b}_{i}\,\Psi = 0\,,\qquad 2\sqrt{2}\,\bar{\cal D}^{ia}{\cal V}^{jk}=
    -\,\varepsilon^{i(j}{\cal D}^{k)a}\,\Psi\,,\qquad
    2\sqrt{2}\,{\cal D}^{ia}{\cal V}^{jk}= -\,\varepsilon^{i(j}\bar{\cal D}^{k)a}\,\bar{\Psi}\,,\nn
    &&\bar{\cal D}^{(i}_a {\cal V}^{jk)}=0\,,\qquad {\cal D}^{(i}_a {\cal V}^{jk)}=0\,,\qquad \tilde{C}\,{\cal V}^{ij}= 0\,,
    \qquad {\cal V}^{ij}\big|_{\vartheta = \bar\vartheta = 0}=v^{ij},
\eea
where ${\cal V}^{ij}$ is an additional deformed ${\cal N}{=}\,8$ superfield.
Solving the constraints, we find that
\bea
    &&A_{ij}=\sqrt{2}\left(-\,i\dot{v}_{ij}+m\,v_{ij}\right),\qquad \pi_{ia}=-\,i\dot{\bar\psi}_{ia}+m\,\bar{\psi}_{ia}-m\,\psi_{ia}\,,\qquad
    D=\ddot{\bar z}+im\dot{\bar z}\,,\nn
    &&\overline{\left(z\right)}=\bar{z}\,,\qquad \overline{\left(\psi_{ia}\right)}=\bar{\psi}^{ia},\qquad
    \overline{\left(v_{ij}\right)}=v^{ij},\qquad \overline{\left(B_{ab}\right)}=B^{ab}=B^{ba}.
\eea
This field content now corresponds to the multiplet ${\bf (5,8,3)}$, and
the deformed transformations \eqref{ch_tr} are rewritten for the involved fields as
\bea
    &&\delta z = -\,\sqrt{2}\,\eta^{ia}\psi_{ia}\,,\qquad
    \delta \bar{z} = \sqrt{2}\,\bar{\eta}^{ia}\bar{\psi}_{ia}\,,\qquad
    \delta v_{ij} = -\,\eta^a_{(j}\bar{\psi}_{i)a} + \bar{\eta}^a_{(j}\psi_{i)a}\,,\nn
    &&\delta \psi_{ia} = 2i\eta^j_a  \dot{v}_{ij}-2m\,\eta^j_a v_{ij} -\sqrt{2}\, \eta^b_i B_{ab} +  \sqrt{2}\,i\bar{\eta}_{ia} \dot{z}\,,\nn
    &&\delta \bar{\psi}_{ia} = -\,2i\bar{\eta}^j_a  \dot{v}_{ij}-2m\,\bar{\eta}^j_a v_{ij} -\sqrt{2}\,\bar{\eta}^b_i B_{ab} -\sqrt{2}\, i\eta_{ia} \dot{\bar{z}}\,,\nn
    &&\delta B_{ab} = \sqrt{2}\,\eta_{i(b}\left[i\dot{\bar\psi}^{i}_{a)}-m\,\bar{\psi}^{i}_{a)}\right]
    + \sqrt{2}\,\bar{\eta}_{i(b}\left[i\dot{\psi}^{i}_{a)}+ m\,\psi^{i}_{a)}\right].\label{583tr}
\eea
\subsection{Invariant actions}
The ${\cal N}{=}\,8$ invariant deformed action can be written as an integral over chiral subspaces, like in the case of
flat ${\cal N}{=}\,8$ supersymmetry \cite{286}:
\bea
    S_{\bf (5,8,3)}=\frac{1}{4}\int d\zeta_{L}\,f\left(\Psi\right) + \frac{1}{4}\int d\zeta_{R}\,\bar{f}\left(\bar{\Psi}\right)
    = \int dt\,{\cal L}_{\bf (5,8,3)}\,.\label{583act}
\eea
The component Lagrangian reads
\bea
    {\cal L}_{\bf (5,8,3)}&=&g\left[\dot{\bar{z}}\dot{z} + \dot{v}_{ij}\dot{v}^{ij} + \frac{i}{2}\left(\psi_{ia}\dot{\bar{\psi}}^{ia}-\dot{\psi}_{ia}\bar{\psi}^{ia}\right)
    -m\,\psi_{ia}\bar{\psi}^{ia}- m^2 v_{ij}v^{ij}+\frac{1}{2}\,B_{ab}B^{ab}\right]\nn
    && - \,\frac{i}{2}\,mg\left(\dot{\bar{z}}z -\dot{z}\bar{z}\right)+\frac{i}{2}\left(\dot{\bar{z}}g_{\bar z} -\dot{z} g_z\right)\psi_{ia}\bar{\psi}^{ia} - \frac{1}{2}
    \left(g_z\,\psi_{a}^{i}\psi_{ib} + g_{\bar z}\,\bar{\psi}_{a}^{i}\bar{\psi}_{ib}\right)B^{ab}\nn
    && + \,\frac{i}{\sqrt{2}}\left(g_z\,\psi_{ia}\psi_{j}^{a}-g_{\bar z}\,\bar{\psi}_{ia}\bar{\psi}_{j}^{a}\right)\dot{v}^{ij} - \frac{m}{\sqrt{2}}\left(g_z\,\psi_{ia}\psi_{j}^{a}+g_{\bar z}\,\bar{\psi}_{ia}\bar{\psi}_{j}^{a}\right)v^{ij}\nn
    && -\,\frac{1}{12}\left(g_{zz}\,\psi^{ib}\psi^{ja}\psi_{jb}\psi_{ia}+g_{\bar{z}\bar{z}}\,\bar{\psi}^{ib}\bar{\psi}^{ja}\bar{\psi}_{jb}\bar{\psi}_{ia}\right).\label{L583}
\eea
Here, $g$ is a special K\"ahler metric defined as
\bea
    g\left(z,\bar{z}\right)=f^{\prime\prime}\left(z\right)+\bar{f}^{\prime\prime}\left(\bar{z}\right),\qquad
    g_z= \frac{\partial g\left(z, \bar{z}\right)}{\partial z}\,,\quad g_{\bar{z}}= \frac{\partial g\left(z, \bar{z}\right)}{\partial\bar{z}}\,,
    \quad {\rm etc}\,. \label{metricSK}
\eea
As compared to the undeformed case, we observe the appearance of the oscillator-type fermionic ($\sim m$) and bosonic ($\sim m^2$) potential terms,
as well as the internal bosonic WZ term accompanied by some new Yukawa-type couplings.

The simplest free action $S^{\rm free}_{\bf (5,8,3)}$ corresponds to the choice $f\left(\Psi\right)=\Psi^2/4\,$. Its component off-shell Lagrangian reads
\be
    {\cal L}^{\rm free}_{\bf (5,8,3)}=\dot{\bar{z}}\dot{z} + \dot{v}_{ij}\dot{v}^{ij} +\frac{i}{2}\left(\psi_{ia}\dot{\bar{\psi}}^{ia}-\dot{\psi}_{ia}\bar{\psi}^{ia}\right)
    -m\,\psi_{ia}\bar{\psi}^{ia}-\frac{i}{2}\,m\left(\dot{\bar{z}}z-\bar{z}\dot{z}\right)- m^2 v_{ij}v^{ij}+\frac{1}{2}\,B_{ab}B^{ab}.
    \label{L583free}
\ee

In \cite{KP2}, $SU(2|2)$ supersymmetry was shown to underlie   ${\cal N}{=}\,8$ massive quantum mechanics of type I inspired by some
super Yang-Mills theory. One can show that the relevant Lagrangian in the abelian case with $U(1)$ as a gauge symmetry
coincides with the on-shell Lagrangian obtained from \eqref{L583free}. It would be interesting to inquire  to which higher-dimensional system
the general Lagrangian \p{L583} could correspond.

\subsection{Description in terms of $SU(2|1)$ superfields}\label{SU21}
The supergroup $SU(2|2)$ contains as a subgroup the supergroup $SU(2|1)$.
Hence, $SU(2|2)$ supersymmetric mechanics can be equivalently viewed as $SU(2|1)$ supersymmetric mechanics \cite{DSQM,SKO,ISTconf,DHSS}
associated with  a few  irreducible $SU(2|1)$  multiplets forming a given $SU(2|2)$ multiplet.
Here, we deal with the supergroup $SU(2|1)$ defined in \eqref{leftSU2}. The multiplet ${\bf (5,8,3)}$ can be split into $SU(2|1)$ multiplets
as ${\bf (4,4,0)}\oplus{\bf (1,4,3)}$ or ${\bf (2,4,2)}\oplus{\bf (3,4,1)}$.
We have restricted our consideration to the latter option.

The $SU(2|1)$ superspace coordinates are defined in the basis \eqref{complexbasis} as
$\left\lbrace t\,,\vartheta_{i1}\,,\bar{\vartheta}^{i1} \right\rbrace =:\left\lbrace t\,, \theta_{i}\,,\bar{\theta}^{i}\right\rbrace$
and are transformed under $SU(2|1)$ according to
\bea
    \delta\theta_{i} = \epsilon_{i} + 2m\,\bar{\epsilon}^{j}\theta_{j}\theta_{i}\,,\qquad
    \delta\bar{\theta}^{i} = \bar{\epsilon}^{i}- 2m\,\epsilon_{j}\bar{\theta}^{j}\bar{\theta}^{i},\qquad
    \delta t = i\bar{\epsilon}^{i}\theta_{i} + i\epsilon_{i}\bar{\theta}^{i}\,.
\eea
Here the parameters $\epsilon_{i}, \bar{\epsilon}^{i}$ are related to the parameters in \p{c_tr} as
\bea
    \eta_{i1} =: \epsilon_{i}\,,\qquad \bar{\eta}^{i1} =: \bar{\epsilon}^{i},\qquad\eta_{i2}
    =: \varepsilon_{i}\,,\qquad \bar{\eta}^{i2} =: \bar{\varepsilon}^{i}\,,
\eea
the $\varepsilon$-transformations being associated with the hidden supersymmetry which extends $SU(2|1)$ to $SU(2|2)$.

The $\epsilon$-transformations in \eqref{583tr} are split into $SU(2|1)$ transformations corresponding
to the chiral multiplet ${\bf (2,4,2)}$ \cite{DSQM} with the $U(1)$ charge $\kappa = 0$ and the multiplet ${\bf (3,4,1)}$ (see Appendix \ref{AppC}):
\bea
    \Phi = \left(z,\psi^{i1},-B_{22}\right)\Rightarrow\left(z,\xi^{i},B\right),\qquad V_{ij}
    =\left(v_{ij},-\bar{\psi}^{i1},-\psi_{i1},-\sqrt{2}\,i B_{12}\right)\Rightarrow\left(v_{ij},\chi^{i},\bar{\chi}_{i},A\right).
\eea
Generally, the $SU(2|2)$ invariant Lagrangian can be written in terms of these $SU(2|1)$ superfields as
\bea
    {\cal L}_{\bf (5,8,3)} = \int d^2\theta\,d^2\bar{\theta}\left(1+2m\,\bar{\theta}^k\theta_k\right){\cal F}\left(\Phi, \bar{\Phi}, V_{ij}\right),\label{su21L}
\eea
where ${\cal F}$ is an arbitrary real scalar function of $SU(2|1)$ superfields satisfying the five-dimensional Laplace equation \cite{Diac,IvSmi}:
\bea
    \left(\frac{4\,\partial^2}{\partial\Phi\,\partial{\bar \Phi}}+\frac{\partial^2}{\partial V^{ij}\,\partial V_{ij}}\right){\cal F}=0\,.\label{583eqs}
\eea
The metric $g:=g\left(z,\bar{z},v_{ij}\right)$ of the target space is expressed as
\bea
    g\left(z, \bar{z}, v_{ij}\right) = -\,\frac{\partial^2{\cal F}\left(z, \bar{z}, v_{ij}\right)}{\partial v^{ij}\,\partial v_{ij}}
    = \frac{4\,\partial^2{\cal F}\left(z, \bar{z}, v_{ij}\right)}{\partial z\,\partial\bar{z}}\,.\label{metric}
\eea
One can explicitly check that \p{583eqs} is the only condition which is required for the invariance under the second subgroup $SU(2)_{\rm R}$ of $SU(2|2)$
in the terms quadratic and quartic in fermions.
Since the closure of $SU(2|1)$ and $SU(2)_{\rm R}$ transformations necessarily yields the supersymmetry $SU(2|2)$,
the equations \eqref{583eqs} is none other than the conditions of the $SU(2|2)$ supersymmetry.
One can treat the invariant Lagrangian \eqref{su21L} as a Lagrangian constructed in terms of harmonic superfields
associated with $\Psi$, $\bar{\Psi}$ and ${\cal V}^{ij}$. This way of obtaining \eqref{su21L} can presumably be figured out from the harmonic formalism
elaborated in \cite{IvSmi}.

As a solution of \eqref{583eqs}, the Lagrangian \eqref{L583} can be rewritten in terms of $SU(2|1)$ superfields as
\bea
    {\cal L}_{\bf (5,8,3)} = \int d^2\theta\,d^2\bar{\theta}\left(1+2m\,\bar{\theta}^k\theta_k\right)
    \left\lbrace\frac{1}{4}\left[\bar{\Phi}f^{\prime}\left(\Phi\right)+\Phi\bar{f}^{\prime}\left(\bar{\Phi}\right)\right]
    -\frac{1}{6}\,V_{ij}V^{ij}\left[f^{\prime\prime}\left(\Phi\right)+\bar{f}^{\prime\prime}\left(\bar{\Phi}\right) \right]\right\rbrace .\nn
\eea
Here, the function $f$ is related to \eqref{583act} as
\bea
    f\left(\Phi\right)\equiv f\left(\Psi\right)\left.\right|_{\vartheta_{i2}=\bar{\vartheta}^{i2}=0},
\eea
and the relevant metric \eqref{metric} coincides with that defined in \eqref{metricSK}.

The metric \eqref{metricSK} corresponds to the most general solution of  \eqref{583eqs} for ${\cal F}$ restricted to the $2$-dimensional target space
as ${\cal F} \equiv {\cal F}\left(z,\bar z\right)\,, g\equiv g\left(z,\bar{z}\right)$. One can consider more general solutions involving some extra dependence on
the triplet $v^{ij}$. For instance, the most general solution with $g\equiv g\left(v_{ij}\right)$ yields the Lagrangian
\bea
    {\cal L}^*_{\bf (5,8,3)} = \frac{1}{8}\int d^2\theta\,d^2\bar{\theta}\left(1+2m\,\bar{\theta}^k\theta_k\right)\frac{\bar{\Phi}\Phi
    - 2 V_{ij}V^{ij}}{\sqrt{V_{ij}V^{ij}\,}}\,.
\eea
In the component form it reads
\bea
    {\cal L}^{*}_{\bf (5,8,3)} &=& \frac{1}{2|v|}\,\bigg[\,\dot{\bar{z}}\dot{z}+\dot{v}_{ij}\dot{v}^{ij} +
\frac{i}{2}\left(\psi_{ia}\dot{\bar{\psi}}^{ia}-\dot{\psi}_{ia}\bar{\psi}^{ia}\right)+\frac{1}{2}\,B^{ab}B_{ab}
+\frac{i}{|v|^2}\,\psi^{(i}_a\bar{\psi}^{j)a}v_{ik}\dot{v}^k_j   \nn
    &&+\, \frac{v^{ij}}{2\sqrt{2}\,|v|^2}
\left(2\,\psi^{a}_{i}\bar{\psi}^{b}_{j}\,B_{ab} + i\psi_{ia}\psi^{a}_j\, \dot{\bar{z}} + i\bar{\psi}_{ia}\bar{\psi}^{a}_j\,\dot{z}\right)
- \frac{3v_{(ij}v_{kl)}}{8|v|^4}\,\psi^a_{i}\psi_{ja}\,\bar{\psi}^b_{k}\bar{\psi}_{lb}\nn
    && - \,\frac{m}{2}\,\psi_{ia}\bar{\psi}^{ia}-m^2v_{ij}v^{ij}\,\bigg]\,,
    \label{583prime}
\eea
with
\bea
    g\left(v_{ij}\right)=\frac{1}{2|v|}\,.
\eea
One can explicitly check that this Lagrangian is invariant under \eqref{583tr}.

There can be many other solutions of \eqref{583eqs} depending on all five fields.
An example of such a solution producing a superconformal model is given in Appendix \ref{AppB3}.

\subsection{The free quantum model}
As an example, here we present quantization of the simplest free model corresponding to \eqref{L583free}.
Eliminating auxiliary fields, we obtain
\be
    {\cal L}^{\rm free}_{\bf (5,8,3)}=\dot{\bar{z}}\dot{z} + \dot{v}_{ij}\dot{v}^{ij} +\frac{i}{2}
    \left(\psi_{ia}\dot{\bar{\psi}}^{ia}-\dot{\psi}_{ia}\bar{\psi}^{ia}\right)
    -m\,\psi_{ia}\bar{\psi}^{ia}-\frac{i}{2}\,m\left(\dot{\bar{z}}z-\bar{z}\dot{z}\right)- m^2 v_{ij}v^{ij}.
\ee
After performing Legendre transformations we obtain the canonical Hamiltonian
\bea
     H=\left(p_{\bar{z}} + \frac{i}{2}\,mz\right)\left(p_z - \frac{i}{2}\,m\bar{z}\right)+\frac{p^{ij}p_{ij}}{4}
     + m^2 v^{ij}v_{ij}+m\,\psi^{ia}\bar{\psi}_{ia}\,.
\eea
Other Noether charges are given by
\bea
    && \Pi^{ia} =\sqrt{2}\left(p_z - \frac{i}{2}\,m\bar{z}\right)\psi^{ia} + \left(p^{ik}+2imv^{ik}\right)\bar{\psi}^{a}_{k},\nn
    && \bar{\Pi}_{jb}=\sqrt{2}\left(p_{\bar{z}} + \frac{i}{2}\,mz\right)\bar{\psi}_{jb} - \left(p_{jk} - 2imv_{jk}\right)\psi^{k}_{b},\nn
    && L^{i}_{j} = \psi^{ia}\bar{\psi}_{ja}-\frac{1}{2}\,\delta^i_j\,\psi^{kc}\bar{\psi}_{kc}+ 2iv^{ik}p_{kj} - i\delta^{i}_{j} v^{kl} p_{kl}  \,,\nn
    && R^a_b=\psi^{ka}\bar{\psi}_{kb}-\frac{1}{2}\,\delta^a_b\,\psi^{kc}\bar{\psi}_{kc}\,.
\eea

The Poisson and Dirac brackets are imposed as
\bea
    \left\lbrace z, p_z \right\rbrace =1\,,\qquad \left\lbrace \bar{z}, p_{\bar{z}} \right\rbrace =1\,, \qquad
    \left\lbrace v^{ij}, p_{kl} \right\rbrace = \frac{1}{2}\left(\delta^{i}_{k}\delta^{j}_{l}+\delta^{i}_{l}\delta^{j}_{k}\right) ,
    \qquad \left\lbrace \bar{\psi}^{ia},\psi_{jb}\right\rbrace = -\,i \delta^{a}_{b} \delta^{i}_{j}
\eea
and they are quantized in the standard way
\bea
    \left[z,p_z\right]=i\,, \qquad \left[\bar{z}, p_{\bar{z}} \right] = i\,, \qquad \left[ v^{ij}, p_{kl} \right]
    = \frac{i}{2}\left(\delta^{i}_{k}\delta^{j}_{l}+\delta^{i}_{l}\delta^{j}_{k}\right) ,\qquad \left\lbrace \bar{\psi}^{ia},\psi_{jb}\right\rbrace
    = \delta^{a}_{b}\delta^{i}_{j} .
\eea
We will use the operators
\bea
    \left[\nabla_z ,\bar{\nabla}_{\bar{z}}\right] = m\,, \qquad
    \left[\nabla^{-ij},\nabla^{+}_{kl}\right] = \frac{m}{2}\left(\delta^{i}_{k}\delta^{j}_{l}+\delta^{i}_{l}\delta^{j}_{k}\right),
\eea
where
\bea
    \nabla_z = p_z - \frac{i}{2}\,m \bar{z} ,\qquad   \bar{\nabla}_{\bar{z}}= p_{\bar{z}}+\frac{i}{2}\,m z, \qquad
    \nabla^{\pm}_{ij} = \frac{1}{2}\left(p_{ij} \pm 2im v_{ij}\right).\label{nabla}
\eea
In terms of the so defined creation and annihilation operators, the quantum version of the generators of \eqref{algebra1} takes the form
\bea
    && \Pi^{ia} =\sqrt{2}\,\nabla_z\psi^{ia} + 2\nabla^{+ik}\bar{\psi}^{a}_{k},\qquad
    \bar{\Pi}_{jb}=\sqrt{2}\,\bar{\nabla}_{\bar{z}}\bar{\psi}_{jb} - 2\nabla^{-}_{jk}\psi^{k}_{b},\nn
    && L^{i}_{j} = \psi^{ia}\bar{\psi}_{ja}-\frac{1}{2}\,\delta^i_j\,\psi^{kc}\bar{\psi}_{kc}+ \frac{1}{m}\left(\nabla^{+ik}\nabla^{-}_{jk}
    - \nabla^{+}_{jk}\nabla^{-ik}\right),\qquad R^a_b=\psi^{ka}\bar{\psi}_{kb}-\frac{1}{2}\,\delta^a_b\,\psi^{kc}\bar{\psi}_{kc}\,,\nn
    && H = \bar{\nabla}_{\bar{z}}\nabla_z + \nabla^{+ij}\nabla^{-}_{ij} + m\,\psi^{ia}\bar{\psi}_{ia}\,.\label{qgen}
\eea
As follows from the definition \eqref{nabla}, the quantum generator $L^{i}_{j}$ in fact does not involve the parameter $m$. So the latter
appears only in the supercharges and the Hamiltonian.

To construct the Hilbert space of wave functions, we use the creation operators $\bar{\nabla}_{\bar{z}}$, $\nabla^{+ij}$, $\psi^{ia}$
and the annihilation operators $\nabla_z$, $\nabla^{-ij}$, $\bar{\psi}^{ia}$. Then, the energy spectrum of $H$
is found to be
\bea
    H\Omega^{(\ell)} = m\ell\,\Omega^{(\ell)},
\eea
where $\Omega^{(\ell)}$ is a wave function at the Landau level $\ell$.
The ground state corresponds to $\ell = 0$ and the first excited level to $\ell=1$.  The relevant wave functions are given by the expressions:
\bea
    &&\Omega^{(0)} = a^{(0)}\left(\bar{z}\right)e^{-\frac{mz\bar{z}}{2}} ,\nn
    &&\Omega^{(1)} = \left(a^{(1)}\left(\bar{z}\right)\bar{\nabla}_{\bar{z}}  + b^{(1)}_{ij}\left(\bar{z}\right)\nabla^{+ij}
    + c^{(1)}_{ia}\left(\bar{z}\right)\psi^{ia}\right) e^{-\frac{mz\bar{z}}{2}}.
\eea
The coefficients $a^{(0)}$, $a^{(1)}$, $b^{(1)}_{ij}$ and $c^{(1)}_{ia}$ are some arbitrary antiholomorphic functions.
This infinite degeneracy is caused by action of the additional generators $\nabla_z + i m \bar{z}$ and $\bar{\nabla} - i m z$ (magnetic translations)
which commute with all quantum generators \eqref{qgen}.
All the higher levels $\ell > 1$ have  wave functions of more complex structure and we will not consider them here.

A few words about $SU(2|2)$ representations are to the point. The ground state $\Omega^{(0)}$
is annihilated by all quantum generators \eqref{qgen}, {\it i.e.}, it is just a singlet. According to \cite{Beisert}, the level $\ell=1$
corresponds to the atypical $SU(2|2)$ representation $\left\langle 1, 0; 1/2,0,0\right\rangle\,,$ with the overall dimension $8$. All
the higher $\ell$ wave functions can be also classified based on the analysis of~\cite{Beisert}.

\setcounter{equation}{0}
\section{Conclusions}
Using powerful $d{=}1$ superfield coset techniques, we have constructed and studied
several models of $SU(2|2)$ supersymmetric mechanics based on the off-shell multiplets
${\bf (3,8,5)}$, ${\bf (4,8,4)}$ and ${\bf (5,8,3)}$.
This new kind of supersymmetric mechanics is a deformation of
flat ${\cal N}{=}\,8$ supersymmetric mechanics.
The corresponding actions were presented, both in terms of superfields and of component fields.
The extended symmetries of these actions were revealed,
and quantization was explicitly performed in one simple case.

Off-shell supermultiplets of standard ${\cal N}{=}\,8$ supersymmetric mechanics
\cite{BIKL1,BIKL2,BHSS,ILS}
other than ${\bf (3,8,5)}$, ${\bf (4,8,4)}$ or ${\bf (5,8,3)}$ do not seem to admit
a deformation to $SU(2|2)$ multiplets.
A possible explanation is the following.
If we take for granted that the $SU(2|2)$ transformations form a subset
of ${\cal N}{=}\,8$ superconformal transformations
(like the $SU(2|1)$ ones which are embedded into an appropriate
${\cal N}{=}\,4$ superconformal group), then such superconformal transformations
can correspond only to the superconformal group $OSp(4^* |4)$.
Indeed, the superalgebra $osp(4^* |4)$ is the only one which contains $su(2|2)$
as a subalgebra (see Appendix~\ref{AppB1}).
According to \cite{KhTo}, $OSp(4^* |4)$ transformations are realized
only on the multiplets ${\bf (3,8,5)}$ and ${\bf (5,8,3)}$.
Hence, the supergroup $SU(2|2)$ also admits an action only on these two multiplets.
The multiplet ${\bf (4,8,4)}$ is exceptional:
none of ${\cal N}{=}\,8$, $d{=}1$ superconformal symmetries can act on it.
However, one can realize on it an ${\cal N}{=}\,8$ extended Heisenberg superalgebra~\cite{BHSS}.
In Section~\ref{484}, we showed that this extended superalgebra \eqref{ext_a1}--\eqref{ext_a2}
contains an $su(2|2)$ superalgebra.
Hence, the $SU(2|2)$ transformations of all three multiplets
${\bf (3,8,5)}$, ${\bf (4,8,4)}$ and ${\bf (5,8,3)}$
are embedded into extended supergroups containing $SU(2|2)$ as a subgroup.

As an example for the contrary, the root multiplet ${\bf (8,8,0)}$ of flat ${\cal N}{=}\,8, d{=}1$
supersymmetry admits $osp(8|2)$ superconformal transformations~\cite{KuTo}.
Besides a flat ${\cal N}{=}\,8, d{=}1$ subalgebra, this superalgebra possesses
two ``curved'' subalgebras with $8$ supercharges~\cite{Sorba}, namely
$su(1|4)$ and $osp(4|2)$,\footnote{
The superalgebra $osp(4|2)$ is a superconformal ${\cal N}{=}\,4$ algebra \cite{AVP}.}
but not $su(2|2)\,$.
This is evidence in favor of the non-existence of the multiplet ${\bf (8,8,0)}$
for $SU(2|2)$ supersymmetry.
Still, one might hope to construct the root multiplet ${\bf (8,8,0)}$ as a sum of
two mutually mirror $SU(2|1)$ multiplets ${\bf (4,4,0)}\oplus {\bf (4,4,0)}$~\cite{DHSS}.
However, by trial and error, we became convinced
that there is no way to extend $SU(2|1)$ supersymmetry to $SU(2|2)$ in such a system,
although we are not able to give a direct rigorous proof.
Similar arguments also suggest the absence of $SU(2|2)$ analogs
of the other ``flat'' off-shell ${\cal N}{=}\,8, d{=}1$ multiplets discussed in~\cite{BIKL2,ILS}.
For a more systematic search of these ``missing''  deformed multiplets,
one should presumably study general deformations of ${\cal N}{=}\,8$ supersymmetric mechanics.

In \cite{KP2}, two types of ${\cal N}{=}\,8$ massive super Yang-Mills quantum mechanics
provided matrix descriptions of supermembranes.
Type I is based on the supergroup $SU(2|2)$, while type II uses the product
supergroup $SU(2|1)\times SU(2|1)$.
In Section~\ref{583}, we noticed that the type I model of~\cite{KP2} reduced
to the simplest $U(1)$ gauge symmetry corresponds to the free Lagrangian~\eqref{L583free}
of the multiplet ${\bf (5,8,3)}$.
Our superfield approach gives the more general $SU(2|2)$ supersymmetric Lagrangian~\eqref{L583}.
It will be interesting to explicitly consider deformations yielding the supergroups
$SU(2|1)\times SU(2|1)$ \cite{KP2} or $SU(1|4)$ \cite{Motl}.
Further off-shell deformed ${\cal N}{=}\,8$ multiplets and the associated mechanics models
may be constructed in this way. It is especially interesting to inspect worldline realizations
of the supergroup $SU(2|4)$~\cite{DSV2,KP1},
as they should bear a direct relation to the matrix models of~\cite{BMN} (see also~\cite{Motl}).
Such models can be studied directly in an $SU(2|2)$ mechanics language,
proceeding from the fact that $SU(2|2)$ is a subgroup of $SU(2|4)$
and representing the multiplets of the latter as direct sums of the appropriate $SU(2|2)$ multiplets.
For instance, the $SU(2|4)$ on-shell multiplet $({\bf 10, 16})$
can hopefully be organized from two copies of  the $SU(2|2)$ multiplet  $({\bf 5, 8, 3})$.
\section*{Acknowledgements}
The work of E.I. and S.S. was supported by RSCF grant no.~16-12-10306.
A part of this study was performed during their visits
to the Institute of Theoretical Physics of Leibniz University of Hannover
within the Heisenberg-Landau program and the joint DFG project LE~838/12-2.
Both thank the directorate of Leibniz University for kind hospitality.
\\

\begin{appendices}
\section{Details of invariant action for the multiplet ${\bf (3,8,5)}$}
\subsection{Calculation of the superfield action}\label{App. A.1}
The idea is to construct the invariant Lagrangian as a power series in $\hat{V}^{++}$,
\be
    L^{(+4)} \propto  \rho\, \tilde{c}^{++}\hat{V}^{++} + m^2\left(\theta^+\right)^4
    \sum_{n=2}^{\infty}a_n\,\big(c^{--}\hat{V}^{++}\big)^n + \sum_{n=2}^{\infty}b_n\left(c^{--}\right)^{n-2}\big(\hat{V}^{++}\big)^n . \lb{Probe}
\ee
Using the fact that $\hat{V}^{++}$ is transformed inhomogeneously in \eqref{Tranvc}, we require that the variations of the adjacent
terms in the sum cancel each other modulo a total derivative, which will impose strict relations between the coefficients $b_n$ and, finally,
fix the form of the above series. We will properly employ the freedom in normalizing \p{Probe}.

The physical normalization of the kinetic term of the boson triplet fixes $b_2 = 1/2$.
We include the transformation of the integration measure \eqref{TranMeas} into the variations of various terms in the Lagrangian \p{Probe}.
Then such a generalized variation of the first term in the sum in \p{Probe} is reduced to (up to a total harmonic derivative)
\be
    \delta \left[\frac{1}{2}\,\big(\hat{V}^{++}\big)^2\right] = \hat{V}^{++}\left[-\,\Lambda \left(\hat{V}^{++} + 2\tilde{c}^{++}\right)
    - 4\Lambda^{++}c^{+-}\right]=-\left(\hat{V}^{++} + 6\tilde{c}^{++}\right)\Lambda\hat{V}^{++}.\lb{varL2}
\ee
The first piece of  this variation is going to be compensated from the variation of the term $\sim \big(\hat{V}^{++}\big)^3$ in \p{Probe},
while the second piece is canceled with the variation of the term $\sim \rho$ in \p{Probe}. Indeed, it is easy to
show that, up to  a total derivative,
\be
    \delta\,\big(\rho\,\tilde{c}^{++}\hat{V}^{++}\big) = 4\Lambda\,\big(\rho\,\tilde{c}^{++}\hat{V}^{++}\big)\,,
\ee
and the choice $\rho = 3/2$
ensures the cancelation needed. Next, we consider the cubic term $b_3\,c^{--}\big(\hat{V}^{++}\big)^3$ with the variation
\be
     \delta\left[b_3\,c^{--}\big(\hat{V}^{++}\big)^3\right] = -\, 4\Lambda\left[b_3\,c^{--}\big(\hat{V}^{++}\big)^3\right]
     - 2b_3\,\Lambda\,\big(\hat{V}^{++}\big)^2 + \frac{5m^2}{2}\,b_3\,\big(c^{--}\hat{V}^{++}\big)^2 \,\delta \left(\theta^+\right)^4. \lb{varL3}
\ee
Under the choice
$b_3 = -\,b_2 = -\,1/2$,
the second piece in \p{varL3} exactly cancels the second piece in \p{varL2}. The remaining term in \p{varL3} is canceled
by the variation of the term
\bea
\frac{5m^2}{4} \left(\theta^+\right)^4 \big(c^{--}\hat{V}^{++}\big)^2 \quad\Rightarrow\quad
    a_2 =  \frac{5}{4}\,.
\eea
In this term, the variation of $\hat{V}^{++}$ yields a vanishing contribution due to the presence of the highest-order
$\theta$ monomial and the fact that both $\Lambda$ and $\Lambda^{++}$ involve at least one power of the Grassmann-odd coordinates. So
its full variation is exclusively defined by the variations of the explicit $\theta$ s.  Proceeding
further, we find that
\bea
&& b_4 = \frac{5}{8}\,, \quad b_5 = -\,\frac{7}{8}\,, \quad {\rm etc}\,, \nn
&& a_3 =  -\,\frac{7}{4}\,, \quad a_4 = \frac{45}{16}\,,
\quad {\rm etc}\,.
\eea
After some effort, using the general formula
\be
    \left(c^{--}\right)^n c^{++} = \frac{n}{2n +1}\,\left(c^{--}\right)^{n-1} + \frac{1}{2\left(n+1\right)
    \left(2n+1\right)}\left({\cal D}^{++}\right)^2 \left(c^{--}\right)^{n+1}\,, \quad n \geqslant 1\,,\quad c^{ij}c_{ij} =1\,,
\ee
we find a recurrence relation for the coefficients $a_n$ and $b_n$\,:
\be
    a_n= -\,\frac{\left(n+1\right)\left(2n+1\right)}{2\left(2n-1\right)}\,b_{n+1}\,,\qquad b_{n+1} = -\,\frac{2n -1}{n+1}\,b_n\,.
\ee
Then, using the property that, up to a total derivative,
\be
\tilde{c}^{++}\hat{V}^{++} = -\,m^2 \left(\theta^+\right)^4 c^{--} \hat{V}^{++},
\ee
we represent the total Lagrangian $L^{(+4)}$ as
\bea
    L^{(+4)} &=& 4\sum_{n=2}^{\infty} (-1)^n\,\frac{\left(2n-3\right)!}{2^n n!\left(n-2\right)!}\,(c^{--})^{n-2}\big(\hat{V}^{++}\big)^n\nn
    && -\,2 m^2 \left(\theta^+\right)^4\,\sum_{n=2}^{\infty}(-1)^n
    \frac{\left(2n-1\right)\left(2n -4\right)!}{ 2^n \left(n-1\right)! \left(n-2\right)!}\,\big(c^{--}\hat{V}^{++}\big)^{n-1}. \label{FullL}
\eea
It remains to learn to which functions these series sum up. Using the Taylor expansions
\bea
&&\sum_{n=0}^\infty \frac{\left(2n+1\right)!}{n!\left(n+2\right)!}\,x^{n+2} = \frac{1}{4}\left(1 - 2x - \sqrt{1 - 4x}\,\right), \nn
&& \sum_{n=0}^\infty \frac{\left(2n\right)!}{\left(n!\right)^2}\,x^{n} = \frac{1}{\sqrt{1 - 4x}\,}\,, \quad
\sum_{n=0}^\infty \frac{\left(2n\right)!}{n!\left(n+1\right)!}\,x^{n+1}
= \frac{1}{2}\left(1 - \sqrt{1 - 4x}\,\right)
\eea
with
$$x := -\,\frac{\hat{V}^{++}c^{--}}{2}\,,$$
it is straightforward to write \eqref{FullL} as
\bea
    L^{(+4)} = \frac{2\,\big(\hat{V}^{++}\big)^2}{\left(1 + \sqrt{1 + 2\,c^{--}\hat{V}^{++}}\,\right)^2} -m^2 \left(\theta^+\right)^4
    \left(\frac{c^{--}\hat{V}^{++}}{\sqrt{1 + 2\,c^{--}\hat{V}^{++}}\,} + \frac{c^{--}\hat{V}^{++}}{1 + \sqrt{1 + 2\,c^{--}\hat{V}^{++}}\,}\right).\nn
    \label{L+4}
\eea
Then, the $SU(2|2)$ invariant Lagrangian for the multiplet ${\bf (3,8,5)}$ is given by
\bea
    {\cal L}_{\bf (3,8,5)}=\int d\zeta^{(-4)}_A L^{(+4)}. \label{L385A}
\eea

\subsection{Harmonic integrals}\label{App. A.2}
For the calculation of the component Lagrangian of \eqref{L385A},
we take as input the known harmonic integrals \cite{GIO, IvLe}
\bea
    &&\int \frac{du }{\left(1 + 2\,c^{--}\hat{v}^{++}\right)^{3/2}} = \frac{1}{\sqrt{1+ 2\,c_{ij}\hat{v}^{ij} + \hat{v}_{ij}\hat{v}^{ij}}\,}\,,\label{int1}\\
    &&\int \frac{du \,u^{+}_{(i}u^{-}_{j)}}{\left(1 + 2\,c^{--}\hat{v}^{++}\right)^{3/2}} = \frac{-\left(c^k_i\hat{v}_{jk}+c^k_j\hat{v}_{ik}\right)}
    {1+2\,c_{ij}\hat{v}^{ij} + \hat{v}_{ij}\hat{v}^{ij} + \left(1 + c_{ij}\hat{v}^{ij}\right)\sqrt{1 + 2\,c_{ij}\hat{v}^{ij} + \hat{v}_{ij}\hat{v}^{ij}}\,}\,.
\eea
After some algebraic manipulations involving integration by parts, the component Lagrangian is reduced to a few terms containing the expressions (all taken at $\theta =0$)
\bea
&& \frac{\partial^2 L^{(+4)}}{\partial \hat{v}^{++}{}^2} = \frac{1}{(1 + 2\,c^{--}\hat{v}^{++})^{3/2}}\,, \nn
&& \frac{\partial^3 L^{(+4)}}{\partial \hat{v}^{++}{}^3} = -\,\frac{3\,c^{--}}{(1 + 2\,c^{--}\hat{v}^{++})^{5/2}}\,, \nn
&& \frac{\partial^4 L^{(+4)}}{\partial \hat{v}^{++}{}^4} = \frac{15\left(c^{--}\right)^2}{(1 + 2\,c^{--}\hat{v}^{++})^{7/2}}\,. \lb{Expr-tu}
\eea
They appear with the specific combinations of harmonics and the corresponding harmonic integrals are computed as
\bea
&& \int du\,\frac{\partial^2 L^{(+4)}}{\partial \hat{v}^{++}{}^2} = \frac{1}{\sqrt{1+ 2\,c_{ij}\hat{v}^{ij} + \hat{v}_{ij}\hat{v}^{ij}}\,}\,, \nn
&& \int du\,u^{+}_k u^{+}_l\,\frac{\partial^3 L^{(+4)}}{\partial \hat{v}^{++}{}^3} = -\,\frac{c_{kl}+\hat{v}_{kl}}{\left(1+ 2\,c_{ij}\hat{v}^{ij} + \hat{v}_{ij}\hat{v}^{ij}\right)^{3/2}}\,, \nn
&& \int du\, u^{+}_k u^{+}_lu^{+}_m u^{+}_n\,\frac{\partial^4 L^{(+4)}}{\partial \hat{v}^{++}{}^4}
=\frac{3\left(c+\hat{v}\right)_{(kl}\left(c+\hat{v}\right)_{mn)}}{\left(1+ 2\,c_{ij}\hat{v}^{ij} + \hat{v}_{ij}\hat{v}^{ij}\right)^{5/2}}\,.
\eea

There is also the harmonic integral
\be
    -\,\int du\left[\frac{\hat{v}^{++}\hat{v}^{--}}{\sqrt{1 + 2 \,c^{--}\hat{v}^{++}}\left(1 + \sqrt{1 + 2\,c^{--}\hat{v}^{++}}\,\right)}+\frac{c^{--}\hat{v}^{++}}{1 + \sqrt{1 + 2\,c^{--}\hat{v}^{++}}\,}+\frac{c^{--}\hat{v}^{++}}{\sqrt{1 + 2 \,c^{--}\hat{v}^{++}}\,}\right],\label{intpot}
\ee
responsible for the potential term $\sim m^2$. It is not immediately obvious how to compute it.
It is easier to calculate this integral by considering its series expansion
\bea
    \int du\sum_{n=2}^{\infty} \,\frac{\left(-1\right)^{n-1}\left(2n-4\right)!}{2^{n-1} \left(n-1\right)!\left(n-2\right)!}\left[\left(2n-3\right)\hat{v}^{--}\hat{v}^{++}+\left(2n-1\right)c^{--}\hat{v}^{++}\right]\left(c^{--}\hat{v}^{++}\right)^{n-2}.\label{series}
\eea
Using the identities
\bea
    &&\int du\left(c^{--}\hat{v}^{++}\right)^{n}
    =\frac{2n-1}{2n+1}\int du\,c_{ij}\hat{v}^{ij}\left(c^{--}\hat{v}^{++}\right)^{n-1}-\frac{n-1}{2n+1}\int du\,\hat{v}^{++}\hat{v}^{--}\left(c^{--}\hat{v}^{++}\right)^{n-2},\nn
    &&\int du\,\hat{v}^{--}\hat{v}^{++}\left(c^{--}\hat{v}^{++}\right)^{n-2}=\frac{\left(n-1\right)\hat{v}_{ij}\hat{v}^{ij}}{2n-1}\int du\left(c^{--}\hat{v}^{++}\right)^{n-2},\nn
    &&\hat{v}_{ij}\hat{v}^{ij}=2\left(\hat{v}^{--}\hat{v}^{++}-\hat{v}^{+-}\hat{v}^{+-}\right),\qquad c_{ij}\hat{v}^{ij}=c^{--}\hat{v}^{++}+c^{++}\hat{v}^{--}-2c^{+-}\hat{v}^{+-},
\eea
we transform \eqref{series} to the form
\bea
    \frac{1}{2}+\left(1+2\,c_{ij}\hat{v}^{ij}+\hat{v}_{ij}\hat{v}^{ij}\right)\int du\sum_{n=0}^{\infty} \,\frac{\left(-1\right)^{n+1}\left(2n+2\right)!}{2^{n+2}n!\left(n+1\right)!}\left(c^{--}\hat{v}^{++}\right)^{n}.
\eea
The final result is given by the integral \eqref{int1} as
\bea
    \eqref{intpot} = -\int \frac{du\left(1+2\,c_{ij}\hat{v}^{ij}+\hat{v}_{ij}\hat{v}^{ij}\right)}{2\left(1 + 2\,c^{--}\hat{v}^{++}\right)^{3/2}}+\frac{1}{2}=\frac{1}{2}\left(1-\sqrt{1+2\,c_{ij}\hat{v}^{ij}+\hat{v}_{ij}\hat{v}^{ij}}\,\right).
\eea

\section{Superconformal symmetry}
Superconformal ${\cal N}{=}\,4$ symmetry in $SU(2|1)$ superspace is realized by {\it trigonometric} transformations \cite{ISTconf}.
Analogously, superconformal ${\cal N}{=}\,8$ symmetry in $SU(2|2)$ superspace also yields trigonometric superconformal mechanics \cite{HT}.

\subsection{Superconformal algebra $osp(4^* |4)$ and its $su(2|2)$ subalgebra}\label{AppB1}
There are four superconformal ${\cal N}{=}\,8$, $d{=}1$ algebras \cite{AVP}: $osp(8|2)$, $su(1,1|4)$, $F(4)$, $osp(4^* |4)$.
According to \cite{Sorba} (Table VI), the superalgebra $su(2|2)$ can be embedded only into the superconformal algebra $osp(4^* |4)$.
Here, we present this embedding.

The explicit structure of $osp(4^* |4)$ is given by the following nonvanishing (anti)commutators:
\bea
    \label{ff}
    && \left\{ Q^{ia},Q^{jb}\right\}=-\,2\,\varepsilon^{ij}\varepsilon^{ab}P,\qquad
 \left\{ {\cal Q}^{i\alpha},{\cal Q}^{j\beta}\right\}=-\,2\,\varepsilon^{ij}\varepsilon^{\alpha\beta}P,\nn
    && \left\{ S^{ia},S^{jb}\right\}=-\,2\,\varepsilon^{ij}\varepsilon^{ab}K,\qquad
 \left\{ {\cal S}^{i\alpha},{\cal S}^{j\beta}\right\}=-\,2\,\varepsilon^{ij}\varepsilon^{\alpha\beta}K,\nn
    && \left\{ Q^{ia},S^{jb}\right\}=2\left(\varepsilon^{ab}T^{ij}-\varepsilon^{ij}\varepsilon^{ab}D-
     2\,\varepsilon^{ij}T_1^{ab}\right),\nn
    && \left\{ {\cal Q}^{i\alpha},{\cal S}^{j\beta}\right\}=2\left(\varepsilon^{\alpha\beta}T^{ij}-
  \varepsilon^{ij}\varepsilon^{\alpha\beta}D-2\,\varepsilon^{ij}T_2^{\alpha\beta}\right),\nn
    && \left\{ Q^{ia},{\cal S}^{j\alpha}\right\}=-\,2\,\varepsilon^{ij}U^{a\alpha},\qquad
 \left\{ {\cal Q}^{i\alpha},S^{ja}\right\}=-\,2\,\varepsilon^{ij}U^{a\alpha},
\eea
\bea
    \label{bb}
    && \left[ D, P\right]=-\,iP,\qquad \left[D, K\right]=iK,\qquad \left[ P, K\right]=2iD, \nn
    && \left[ T^{ij},T^{kl}\right]= -\,i\left(\varepsilon^{ik}T^{jl}+\varepsilon^{jl}T^{ik}\right),\qquad
    \left[U^{a\alpha},U^{b\beta}\right]=
-\,2i\left(\varepsilon^{\alpha\beta}T_1^{ab}+\varepsilon^{ab}T_2^{\alpha\beta}\right),\nn
    && \left[ T_1^{ab},T_1^{cd}\right]=-\,i\left(\varepsilon^{ac}T_1^{bd}+\varepsilon^{bd}T_1^{ac}\right),\qquad
  \left[ T_2^{\alpha\beta},T_2^{\gamma\rho}\right]=-\,i\left(\varepsilon^{\alpha\gamma}T_2^{\beta\rho}+
 \varepsilon^{\beta\rho}T_2^{\alpha\gamma}\right),\nn
    && \left[T_1^{ab},U^{c\alpha}\right]=-\,\frac{i}{2}\left( \varepsilon^{ac}U^{b\alpha}+\varepsilon^{bc}U^{a\alpha}\right),\qquad
    \left[T_2^{\alpha\beta},U^{a\gamma}\right]=-\,\frac{i}{2}\left( \varepsilon^{\alpha\gamma}U^{a\beta}+
 \varepsilon^{\beta\gamma}U^{a\alpha}\right),
\eea
\bea
    \label{bf}
    &&\left[P, S^{ia}\right]=i\,Q^{ia},\quad
    \left[P, {\cal S}^{i\alpha}\right]=i\,{\cal Q}^{i\alpha},\quad
    \left[K, Q^{ia}\right]=-\,i\, S^{ia},\quad \left[K, {\cal Q}^{i\alpha}\right]=-\,i\,{\cal S}^{i\alpha},\nn
    && \left[D, Q^{ia}\right]=-\,\frac{i}{2}\,Q^{ia},\quad \left[D, {\cal Q}^{i\alpha}\right]=-\,\frac{i}{2}\,{\cal Q}^{i\alpha},\quad
    \left[D, S^{ia}\right]=\frac{i}{2}\,S^{ia},\quad \left[D, {\cal S}^{i\alpha}\right]=\frac{i}{2}\,{\cal S}^{i\alpha},\nn
    && \left[ U^{a\alpha},Q^{ib}\right]= -\,i\,\varepsilon^{ab}{\cal Q}^{i\alpha},\quad
    \left[ U^{a\alpha},{\cal Q}^{i\beta}\right]= -\,i\,\varepsilon^{\alpha\beta}Q^{ia},\nn
    && \left[ U^{a\alpha},S^{ib}\right]= -\,i\,\varepsilon^{ab}{\cal S}^{i\alpha},\quad
    \left[ U^{a\alpha},{\cal S}^{i\beta}\right]= -\,i\,\varepsilon^{\alpha\beta}S^{ia},\nn
    && \left[T_1^{ab},Q^{ic}\right]=-\,\frac{i}{2}\left( \varepsilon^{ac}Q^{ib}+\varepsilon^{bc}Q^{ia}\right),\qquad
    \left[T_1^{ab},S^{ic}\right]=-\,\frac{i}{2}\left( \varepsilon^{ac}S^{ib}+\varepsilon^{bc}S^{ia}\right),\nn
    && \left[T_2^{\alpha\beta},{\cal Q}^{i\gamma}\right]=-\,\frac{i}{2}\left(\varepsilon^{\alpha\gamma}{\cal Q}^{i\beta}+\varepsilon^{\beta\gamma}{\cal Q}^{i\alpha}\right),\qquad
    \left[T_2^{\alpha\beta},{\cal S}^{i\gamma}\right]=-\,\frac{i}{2}\left(\varepsilon^{\alpha\gamma}{\cal S}^{i\beta}+\varepsilon^{\beta\gamma}{\cal S}^{i\alpha}\right),\nn
    && \left[T^{ij},Q^{ka}\right]=-\,\frac{i}{2}\left( \varepsilon^{ik}Q^{ja}+\varepsilon^{jk}Q^{ia}\right),\qquad
    \left[T_1^{ij},S^{ka}\right]=-\,\frac{i}{2}\left( \varepsilon^{ik}S^{jb}+\varepsilon^{jk}S^{ia}\right),\nn
    && \left[T^{ij},{\cal Q}^{k\alpha}\right]=-\,\frac{i}{2}\left(\varepsilon^{ik}{\cal Q}^{j\alpha}+\varepsilon^{jk}{\cal Q}^{i\alpha}\right),\qquad
    \left[T^{ij},{\cal S}^{k\alpha}\right]=-\,\frac{i}{2}\left(\varepsilon^{ik}{\cal S}^{j\alpha}+\varepsilon^{jk}{\cal S}^{i\alpha}\right).
\eea
The bosonic subgroup is $SO(5)\times SU(2)\times SO(2,1)$.

There are three $SU(2)$ groups with generators $T^{ij}$, $T_1^{ab}$, $T_2^{\alpha\beta}$ acting on the relevant indices.
Let us redefine the $SU(2)$ indices $a,b,c\ldots$ and $\alpha,\beta,\gamma,\ldots$ as
\bea
    {\cal Q}^{i\alpha} \rightarrow {\cal Q}^{ia},\qquad {\cal S}^{i\alpha} \rightarrow {\cal S}^{ia},\qquad T_2^{\alpha\beta}\rightarrow T_2^{ab},
    \qquad
    U^{a\alpha}\rightarrow U^{ab}.
\eea
This redefinition just means that we passed to the equivalent basis where, instead of the $su(2)$ algebra with the generators $T_2^{\alpha\beta}$,
we deal with the diagonal $su(2)$ in the direct sum of the former  $su(2)$ and the one with the generators $T_1^{ab}$.
Then, we define the $m$-deformed supercharges
\bea
    &&\Pi^{ia}(\pm m) :=\frac{1}{2}\left[Q^{ia}-i{\cal Q}^{ia}\pm m\left({\cal S}^{ia} + iS^{ia}\right)\right],\nn
    &&\bar{\Pi}^{ia}(\pm m) := \frac{1}{2}\left[Q^{ia}+i{\cal Q}^{ia}\pm m\left({\cal S}^{ia} - iS^{ia}\right)\right]
\eea
and the bosonic generators
\bea
    && {\cal H}=-\,\frac{1}{2}\left(P+m^2K\right),\qquad U=\frac{1}{2}\,\varepsilon_{cd}\,U^{cd},\qquad U^{(ab)}=\frac{1}{2}\left(U^{ab}+U^{ba}\right),\qquad
    L^{ij} = i\,T^{ij}, \nn
    &&R^{ab} = i\left(T_1^{ab}+T_2^{ab}\right),\qquad \tilde{R}^{ab} = i\left(T_1^{ab}-T_2^{ab}\right),\qquad {\cal T}_{\pm}=\frac{1}{2}\left(P-m^2K\pm 2imD\right).
\eea
In terms of these redefined generators the superalgebra $osp(4^* |4)$ takes the form:
\bea
    &&\left\lbrace \Pi^{ia}(\pm m), \bar{\Pi}^{jb}(\pm m)\right\rbrace = \mp\, 2m\left(\varepsilon^{ab}L^{ij}-\varepsilon^{ij}R^{ab}\right) + 2\,\varepsilon^{ab}\varepsilon^{ij}\left({\cal H}\pm m U\right),\nn
    &&\left\lbrace \Pi^{ia}(\pm m), \bar{\Pi}^{jb}(\mp m)\right\rbrace = -\,\varepsilon^{ij}\varepsilon^{ab}{\cal T}_{\pm}\,,\nn
    &&\left\lbrace \Pi^{ia}(m), \Pi^{jb}(-m)\right\rbrace = 2m\,\varepsilon^{ij}U^{(ab)}+2m\,\varepsilon^{ij}\tilde{R}^{ab},\nn
    &&\left\lbrace \bar{\Pi}^{ia}(m), \bar{\Pi}^{jb}(-m)\right\rbrace = 2m\,\varepsilon^{ij}U^{(ab)}-2m\,\varepsilon^{ij}\tilde{R}^{ab},
\eea
\bea
    && \left[{\cal H},{\cal T}_{\pm}\right]=\pm\,m\,{\cal T}_{\pm}\,,\qquad \left[{\cal T}_{+},{\cal T}_{-}\right]=-\,2m {\cal H},\nn
    && \left[L^{ij}, L^{kl}\right] = \varepsilon^{il}L^{kj} +\varepsilon^{jk}L^{il}, \qquad
    \left[R^{ab}, \tilde{R}^{cd}\right] = \varepsilon^{ad}\tilde{R}^{bc} +\varepsilon^{bc}\tilde{R}^{ad},\nn
    && \left[R^{ab}, R^{cd}\right] = \varepsilon^{ad}R^{bc} +\varepsilon^{bc}R^{ad},\qquad
    \left[\tilde{R}^{ab}, \tilde{R}^{cd}\right] = \varepsilon^{ad}R^{bc} +\varepsilon^{bc}R^{ad},\nn
    && \left[R^{ab}, U^{(cd)}\right] = \varepsilon^{ad}U^{(bc)} +\varepsilon^{bc}U^{(ad)},\qquad
    \left[\tilde{R}^{ab}, U^{(cd)}\right] = -\left(\varepsilon^{ac}\varepsilon^{bd}+\varepsilon^{ad}\varepsilon^{bc}\right)U,\nn
    && \left[U^{(ab)}, U^{(cd)}\right]=-\,\varepsilon^{ad}R^{bc}-\varepsilon^{bc}R^{ad},\qquad \left[\tilde{R}^{ab}, U\right] = U^{(ab)},\qquad
    \left[U^{(ab)}, U\right]=\tilde{R}^{ab},
\eea
\bea
    && \left[L^{ij}, \Pi^{ka}(\pm m)\right] = \frac{1}{2}\left(\varepsilon^{ik}\Pi^{ja}(\pm m) + \varepsilon^{jk}\Pi^{ia}(\pm m)\right), \nn
    && \left[R^{ab}, \Pi^{ic}(\pm m)\right] = \frac{1}{2}\left(\varepsilon^{ac}\Pi^{ib}(\pm m) + \varepsilon^{bc}\Pi^{ia}(\pm m)\right),\nn
    && \left[L^{ij}, \bar{\Pi}^{ka}(\pm m)\right] = \frac{1}{2}\left(\varepsilon^{ik}\bar{\Pi}^{ja}(\pm m) + \varepsilon^{jk}\bar{\Pi}^{ia}(\pm m)\right), \nn
    && \left[R^{ab}, \bar{\Pi}^{ic}(\pm m)\right] = \frac{1}{2}\left(\varepsilon^{ac}\bar{\Pi}^{ib}(\pm m) + \varepsilon^{bc}\bar{\Pi}^{ia}(\pm m)\right),\nn
    && \left[\tilde{R}^{ab}, \Pi^{ic}(\pm m)\right] = \frac{1}{2}\left(\varepsilon^{ac}\bar{\Pi}^{ib}(\mp m) + \varepsilon^{bc}\bar{\Pi}^{ia}(\mp m)\right),\nn
    && \left[\tilde{R}^{ab}, \bar{\Pi}^{ic}(\pm m)\right] = \frac{1}{2}\left(\varepsilon^{ac}\Pi^{ib}(\mp m) + \varepsilon^{bc}\Pi^{ia}(\mp m)\right),
\eea
\bea
    && \left[{\cal H},\Pi^{ia}(\pm m)\right]=\pm\,\frac{m}{2}\,\Pi^{ia}(\pm m),\qquad
    \left[{\cal H},\bar{\Pi}^{ia}(\pm m)\right]=\mp\,\frac{m}{2}\,\bar{\Pi}^{ia}(\pm m),\nn
    && \left[{\cal T}_{\pm},\Pi^{ia}(\mp m)\right]=\pm\,m\,\Pi^{ia}(\pm m),\qquad \left[{\cal T}_{\pm},\bar{\Pi}^{ia}(\mp m)\right]=\mp\,m\,\bar{\Pi}^{ia}(\pm m),\nn
    && \left[U,\Pi^{ia}(\pm m)\right]=-\,\frac{1}{2}\,\Pi^{ia}(\pm m),\qquad
    \left[U,\bar{\Pi}^{ia}(\pm m)\right]=\frac{1}{2}\,\bar{\Pi}^{ia}(\pm m),\nn
    && \left[U^{(ab)}, \Pi^{ic}(\pm m)\right] = -\,\frac{1}{2}\left(\varepsilon^{ac}\bar{\Pi}^{ib}(\mp m) + \varepsilon^{bc}\bar{\Pi}^{ia}(\mp m)\right),\nn
    && \left[U^{(ab)}, \bar{\Pi}^{ic}(\pm m)\right] = \frac{1}{2}\left(\varepsilon^{ac}\Pi^{ib}(\mp m) + \varepsilon^{bc}\Pi^{ia}(\mp m)\right).
\eea
Thus, the deformed supercharges $\Pi^{ia}(m)$, $\bar{\Pi}^{jb}(m)$ generate the $su(2|2)$ superalgebra
\eqref{algebra1} with central charges $H={\cal H}+mU$ and $C=0$.\footnote{As discussed, any ${\cal N}{=}\,8$, $d{=}1$ superconformal symmetry cannot be realized on the multiplet $({\bf 4,8,4})$ which has $C\neq 0$.} Hence, the supercharges
$\Pi^{ia}(-m)$, $\bar{\Pi}^{jb}(-m)$ also form a $su(2|2)$ superalgebra, but with the opposite-sign deformation parameter $-m$.
The closure of these two $su(2|2)$ superalgebras yields the whole superconformal algebra $osp(4^* |4)$.
In this embedding, the subgroup $SO(5)$ contains the subgroup $SU(2)_{\rm R}$ from $SU(2|2)$, {\it i.e.} $SU(2)_{\rm R}\subset SO(5)$. Switching $SU(2)_{\rm R}\leftrightarrow SU(2)_{\rm L}$ in $SU(2|2)$, one can consider the embedding where $SU(2)_{\rm L}\subset SO(5)$.

\subsection{Superconformal properties of the multiplet ${\bf (3,8,5)}$}
The component fields of \eqref{compv++} can be rewritten in a complex notation as
\bea
    &&\xi_{ia}=\frac{1}{\sqrt{2}}\left(\bar{\psi}_{ia}\,e^{-\frac{i}{2}mt}-\psi_{ia}\,e^{\frac{i}{2}mt}\right),\qquad
    \hat{\xi}_{ia}=\frac{i}{\sqrt{2}}\left(\bar{\psi}_{ia}\,e^{-\frac{i}{2}mt}+\psi_{ia}\,e^{\frac{i}{2}mt}\right),\nn
    &&A_0=-\,\frac{i}{2}\left(A\,e^{imt}+\bar{A}\,e^{-imt}\right),\qquad
    C_0=\frac{1}{2}\left(A\,e^{imt}-\bar{A}\,e^{-imt}\right).
\eea
Then the Lagrangian \eqref{KinComp1} is rewritten as
\bea
    {\cal L}^{\rm conf.}_{\bf (3,8,5)} &=& \frac{1}{2|v|}\,\bigg[\,\dot{v}_{ij}\dot{v}^{ij} +
\frac{i}{2}\left(\psi_{ia}\dot{\bar{\psi}}^{ia}-\dot{\psi}_{ia}\bar{\psi}^{ia}\right)
+\frac{i}{|v|^2}\,\psi^{(i}_a\bar{\psi}^{j)a}v_{ik}\dot{v}^k_j +2A\bar{A}-\frac{1}{4}\,C^{ab}C_{ab}  \nn
    &&+\, \frac{v^{ij}}{2|v|^2}
\left(i\psi^{a}_{i}\bar{\psi}^{b}_{j}\,C_{ab} + i\psi_{ia}\psi^{a}_j\, \bar{A} + i\bar{\psi}_{ia}\bar{\psi}^{a}_j\,A
-2\mu\,\psi_{ia}\bar{\psi}^{a}_j\right)
    - \frac{3v_{(ij}v_{kl)}}{8|v|^4}\,\psi^a_{i}\psi_{ja}\,\bar{\psi}^b_{k}\bar{\psi}_{lb}\nn
    &&+\,\mu^2 - m^2v_{ij}v^{ij}\,\bigg]  -\frac{i\mu\,\dot{v}^{ij}\left(c^k_i v_{jk}+c^k_j v_{ik}\right)}
    {|v|\left(|v| + c_{ij} v^{ij}\right)}\,.\label{Lconf}
\eea
The relevant transformations \eqref{TrnsfComp} leaving this Lagrangian invariant (modulo a total derivative), become
\bea
    &&\delta v_{ij} = -\,\eta^a_{(j}\bar{\psi}_{i)a}\,e^{-\frac{i}{2}mt} + \bar{\eta}^a_{(j}\psi_{i)a}\,e^{\frac{i}{2}mt},\nn
    &&\delta \psi_{ia} = \left(2i\eta^j_a  \dot{v}_{ij}-2m\,\eta^j_a v_{ij} -i \eta^b_i C_{ab} +2\mu\,\eta_{ia}\right) e^{-\frac{i}{2}mt}+  2i\bar{\eta}_{ia} A\,e^{\frac{i}{2}mt},\nn
    &&\delta \bar{\psi}_{ia} = -\left(2i\bar{\eta}^j_a  \dot{v}_{ij}+2m\,\bar{\eta}^j_a v_{ij} + i\bar{\eta}^b_i C_{ab}+2\mu\,\bar{\eta}_{ia}\right)e^{\frac{i}{2}mt} - 2i\eta_{ia} \bar{A}\,e^{-\frac{i}{2}mt},\nn
    &&\delta C_{ab} = 2\eta_{i(b}\left[\dot{\bar\psi}^{i}_{a)}+\frac{i}{2}\,m\,\bar{\psi}^{i}_{a)}\right]e^{-\frac{i}{2}mt} + 2\bar{\eta}_{i(b}\left[\dot{\psi}^{i}_{a)} - \frac{i}{2}\,m\,\psi^{i}_{a)}\right]e^{\frac{i}{2}mt},\nn
    &&\delta A = -\,\eta^{ia}\left(\dot{\psi}_{ia}+\frac{i}{2}\,m\,\psi_{ia}\right)e^{-\frac{i}{2}mt},\qquad
    \delta \bar{A} = \bar{\eta}^{ia}\left(\dot{\bar{\psi}}_{ia}-\frac{i}{2}\,m\,\bar{\psi}_{ia}\right)e^{\frac{i}{2}mt}.
    \label{385tr}
\eea
Exploiting the property that the Lagrangian \eqref{Lconf} depends only on $m^2$, we can define additional $SU(2|2)$ transformations with
$m\rightarrow -m$:
\bea
    &&\delta^\prime v_{ij} = -\,\eta^{\prime a}_{(j}\bar{\psi}_{i)a}\,e^{\frac{i}{2}mt} + \bar{\eta}^{\prime a}_{(j}\psi_{i)a}\,e^{-\frac{i}{2}mt},\nn
    &&\delta^\prime \psi_{ia} = \left(2i\eta^{\prime j}_a  \dot{v}_{ij}+2m\,\eta^{\prime j}_a v_{ij} -i \eta^{\prime b}_i C_{ab} +2\mu\,\eta^{\prime}_{ia}\right) e^{\frac{i}{2}mt}+  2i\bar{\eta}^{\prime}_{ia} A\,e^{-\frac{i}{2}mt},\nn
    &&\delta^\prime \bar{\psi}_{ia} = -\left(2i\bar{\eta}^{\prime j}_a  \dot{v}_{ij}-2m\,\bar{\eta}^{\prime j}_a v_{ij} + i\bar{\eta}^{\prime b}_i C_{ab}+2\mu\,\bar{\eta}^\prime_{ia}\right)e^{-\frac{i}{2}mt} - 2i\eta^{\prime}_{ia} \bar{A}\,e^{\frac{i}{2}mt},\nn
    &&\delta^\prime C_{ab} = 2\eta^\prime_{i(b}\left[\dot{\bar\psi}^{i}_{a)} - \frac{i}{2}\,m\,\bar{\psi}^{i}_{a)}\right]e^{\frac{i}{2}mt} + 2\bar{\eta}^\prime_{i(b}\left[\dot{\psi}^{i}_{a)} + \frac{i}{2}\,m\,\psi^{i}_{a)}\right]e^{-\frac{i}{2}mt},\nn
    &&\delta^\prime A = -\,\eta^{\prime ia}\left(\dot{\psi}_{ia}-\frac{i}{2}\,m\,\psi_{ia}\right)e^{\frac{i}{2}mt},\qquad
    \delta^\prime \bar{A} = \bar{\eta}^{\prime ia}\left(\dot{\bar{\psi}}_{ia}+\frac{i}{2}\,m\,\bar{\psi}_{ia}\right)e^{-\frac{i}{2}mt}.
    \label{385conftr}
\eea
The closure of these two types of $SU(2|2)$ transformations gives
a trigonometric realization of the full superconformal symmetry $OSp(4^*|4)$.
Thus, the Lagrangian \eqref{Lconf} is superconformal and is recognized as a deformation of the parabolic Lagrangian given in \cite{BIKL2}
by the oscillator mass term $\sim m^2$.

\subsection{Superconformal properties of the multiplet ${\bf (5,8,3)}$}\label{AppB3}
Redefining component fields in \eqref{583tr} as
\bea
    &&z \rightarrow z e^{-imt},\qquad \bar{z}\rightarrow \bar{z}e^{imt},
    \qquad B_{ab} \rightarrow B_{ab}\,,\nn
    &&\psi_{ia} \rightarrow \psi_{ia}\, e^{-\frac{i}{2}mt},\qquad
    \bar{\psi}_{ia} \rightarrow \bar{\psi}_{ia}\, e^{\frac{i}{2}mt},
\eea
we cast their $SU(2|2)$ transformations into the form
\bea
    &&\delta z = -\,\sqrt{2}\,\eta^{ia}\psi_{ia}\,e^{\frac{i}{2}mt},\qquad
    \delta \bar{z} = \sqrt{2}\,\bar{\eta}^{ia}\bar{\psi}_{ia}\,e^{-\frac{i}{2}mt},\qquad
    \delta v_{ij} = -\,\eta^a_{(j}\bar{\psi}_{i)a}\,e^{\frac{i}{2}mt} + \bar{\eta}^a_{(j}\psi_{i)a}\,e^{-\frac{i}{2}mt},\nn
    &&\delta \psi_{ia} =\left(2i\eta^j_a  \dot{v}_{ij}-2m\,\eta^j_a v_{ij} -\sqrt{2}\, \eta^b_i B_{ab}\right) e^{\frac{i}{2}mt} +  \sqrt{2}\,\bar{\eta}_{ia}\left(i\dot{z}+mz\right)e^{-\frac{i}{2}mt},\nn
    &&\delta \bar{\psi}_{ia} = \left(-\,2i\bar{\eta}^j_a  \dot{v}_{ij}-2m\,\bar{\eta}^j_a v_{ij} -\sqrt{2}\,\bar{\eta}^b_i B_{ab}\right)e^{-\frac{i}{2}mt} -\sqrt{2}\, \eta_{ia} \left(i\dot{\bar{z}}-m\bar{z}\right)e^{\frac{i}{2}mt},\nn
    &&\delta B_{ab} = \sqrt{2}\,\eta_{i(b}\left[i\dot{\bar\psi}^{i}_{a)}-\frac{3m}{2}\,\bar{\psi}^{i}_{a)}\right]e^{\frac{i}{2}mt} + \sqrt{2}\,\bar{\eta}_{i(b}\left[i\dot{\psi}^{i}_{a)}+ \frac{3m}{2}\,\psi^{i}_{a)}\right]e^{-\frac{i}{2}mt}.
\eea
Making the change $m \rightarrow -m$ in these transformations, we define additional $SU(2|2)$ transformations.
 In the same way as in the previous case, the two types of $SU(2|2)$ transformations close on the superconformal symmetry $OSp(4^*|4)$
in the trigonometric realization.

Superconformal Lagrangian admits construction in terms of $SU(2|1)$ superfields corresponding to the multiplets ${\bf (2,4,2)}$ and ${\bf (3,4,1)}$ as $SU(1,1|2)\subset OSp(4^*|4)$ superconformal trigonometric Lagrangian. Superfield Lagrangian satisfying \eqref{583eqs} is given by
\bea
    {\cal L}^{\rm conf.}_{\bf (5,8,3)} = \int d^2\theta\,d^2\bar{\theta}\left(1+2m\,\bar{\theta}^k\theta_k\right)\frac{\log{\left(\sqrt{V_{ij}V^{ij}}+\sqrt{V_{ij}V^{ij}+\Phi\bar{\Phi}}\right)}}{\sqrt{V_{ij}V^{ij}}}.
\eea
In contrast to Sec. \ref{SU21}, the chiral $SU(2|1)$ superfield $\Phi$ describing the multiplet ${\bf (2,4,2)}$ has a central charge~\footnote{The central charge $b$ is related to a scaling dimension parameter as $b=-\lambda_D$ where $\lambda_d{=}1$ for the multiplet ${\bf (5,8,3)}$ \cite{KhTo}. Thus, we have that  $b=-1$.} $b=-1$ \cite{ISTconf}.
By analogy with \eqref{Lconf} and previously constructed trigonometric superconformal Lagrangians \cite{HT,ISTconf}, the relevant ${\bf (5,8,3)}$ superconformal Lagrangian is a deformation of the parabolic superconformal Lagrangian \cite{BIKL1,BIKL2} by oscillator term.
The bosonic truncation of superconformal Lagrangian reads
\bea
    {\cal L}^{\rm conf.}_{\bf (5,8,3)}\left.\right|_{\rm bos} &=& \left(v_{ij}v^{ij}+z\bar{z}\right)^{-3/2}\left[\dot{\bar{z}}\dot{z}+\dot{v}_{ij}\dot{v}^{ij} +
\frac{1}{2}\,B^{ab}B_{ab}-m^2 \left(v_{ij}v^{ij}+z\bar{z}\right)\right].   \label{583conf}
\eea
One can see that this Lagrangian is $SO(5)\times SU(2)$ invariant, where dynamical bosonic fields form $SO(5)$ vector and auxiliary fields are combined into $SU(2)$ triplet.

\section{The multiplet ${\bf (3,4,1)}$}\label{AppC}
We briefly consider the multiplet ${\bf (3,4,1)}$ described by the superfield $V^{ij}$ in the framework of the $SU(2|1)$ superspace \cite{DSQM}.
It satisfies the $SU(2|1)$ covariant constraints
\bea
    {\cal D}^{(k}V^{ij)} = \bar{{\cal D}}^{(k}V^{ij)} =0\,, \qquad \left( V^{ij}\right)^\dagger =V_{ij}\,,\qquad V^{ij}\equiv V^{ji},
\eea
where $SU(2|1)$ covariant derivatives are
\bea
    {\cal D}^i &=& \left[1+{m}\,\bar{\theta}^k\theta_k
    -\frac{3m^2}{4}\left(\bar{\theta}^k\theta_k\right)^2\right]\frac{\partial}{\partial\theta_i}
    - {m}\,\bar{\theta}^i\theta_j\frac{\partial}{\partial\theta_j}-i\bar{\theta}^i \partial_t
    +m\,\bar{\theta}^i \tilde{F}- {m}\,\bar{\theta}^j\left(1 -m\,\bar{\theta}^k\theta_k \right)\tilde{I}^i_j\,,\nn
    \bar{{\cal D}}_j &=& -\left[1+ {m}\,\bar{\theta}^k\theta_k
    -\frac{3m^2}{4}\left(\bar{\theta}^k\theta_k\right)^2\right]\frac{\partial}{\partial\bar{\theta}^j}
    + {m}\,\bar{\theta}^k\theta_j\frac{\partial}{\partial\bar{\theta}^k}+i\theta_j\partial_t
    -m\,\theta_j\tilde{F}+ {m}\,\theta_k\left(1 -m\,\bar{\theta}^l\theta_l \right)\tilde{I}^k_j\,.\nn
\eea
The solution is given by
\bea
    V^{ij}&=&\left[1+ m\,\bar{\theta}^k\theta_k- m^2\left(\bar{\theta}^k\theta_k\right)^2\right]v^{ij}
    + \theta^{(i} \chi^{j)} +\bar{\theta}^{(i}\bar{\chi}^{j)} - i\bar{\theta}^{(i} \theta^{j)} A\nn
    &&- \,i\left[\bar{\theta}^{(i} \theta_k + \bar{\theta}_k \theta^{(i}\right]\dot{v}^{j)k}
    -   i\bar{\theta}^k\theta_k\left[\theta^{(i} \dot{\chi}^{j)}
    -\bar{\theta}^{(i}\dot{\bar{\chi}}^{j)}\right] +\frac{1}{2}\left(\bar{\theta}^k\theta_k\right)^2\ddot{v}^{ij},\nn
    && \overline{\left( v^{ij}\right)} =v_{ij}\,,\quad \overline{\left(\chi^{i}\right)}
    =\bar{\chi}_{i}\,,\quad \overline{\left(B\right)}=B. \label{V2}
\eea
The superfield $V^{ij}$ has the following passive transformations
\bea
    \delta V^{ij} = -\,m\left(1+m\,\bar{\theta}^l\theta_l\right)
    \left[\left(\epsilon_k\bar{\theta}^k + \bar{\epsilon}^k\theta_k\right)V^{ij}-\left(\epsilon_k\bar{\theta}^i
    + \bar{\epsilon}^i\theta_k\right)V^{kj}-\left(\epsilon_k\bar{\theta}^j + \bar{\epsilon}^j\theta_k\right)V^{ik}\right],
\eea
and its component fields transform as
\bea
\label{341tr}
    && \delta v^{ij} =-\,\epsilon^{(i}\chi^{j)} -\bar{\epsilon}^{(i} \bar{\chi}^{j)} ,\qquad
    \delta A= \epsilon_k\dot{\chi}^k-\bar{\epsilon}^k\dot{\bar\chi}_k
    + im\left(\epsilon_k\chi^k +\bar{\epsilon}^k\bar{\chi}_k\right),\nn
    && \delta \bar{\chi}_j =-\,2i\epsilon^k \dot{v}_{kj} +i\epsilon_j A+ 2m\,\epsilon^k v_{kj}\,,\qquad
    \delta\chi^i = -\,2i\bar{\epsilon}_k \dot{v}^{ik}- i\bar{\epsilon}^i A - 2m\,\bar{\epsilon}_k v^{ik}.
\eea

The general $\sigma$-model action for the $SU(2|1)$ multiplet $({\bf 3, 4, 1})$ is constructed as
\bea
    S_{\bf (3,4,1)}=\int dt\,{\cal L} =  -\,\frac{1}{6}\int dt\,d^2\theta\,
    d^2\bar{\theta}\left(1+2m\,\bar{\theta}^k\theta_k\right) L\left(V^{2}\right),\qquad V^2=V_{ij}V^{ij},\label{SV}
\eea
where $dt\,d^2\theta\,d^2\bar{\theta}\left(1+2m\,\bar{\theta}^k\theta_k\right)$ is a $SU(2|1)$ invariant measure.
The simplest free Lagrangian corresponding to the choice $L = V^2$ reads
\bea
    {\cal L}^{\rm free}_{\bf (3,4,1)}=\dot{v}_{ij}\dot{v}^{ij} + \frac{i}{2}\left(\bar{\chi}_i\dot{\chi}^i - \dot{\bar{\chi}}_i\chi^i\right)
    +m\,\chi^i\bar{\chi}_i - m^2 v_{ij} v^{ij}+\frac{A^2}{2}\,.\label{341free}
\eea
After eliminating the auxiliary field $A$, this Lagrangian can be viewed as an abelian reduction of the mass-deformed ${\cal N}{=}\,4$ matrix models
of type I \cite{KP2}. Recently, in \cite{Denef}, the $SU(2|1)$ multiplet $({\bf 3,4,1})$ was used at the component level
for the description of a new class of ${\cal N}{=}\,4$ supersymmetric massive quiver matrix models. The component Lagrangian \eqref{341free}
corresponds to the simplest case of one node without arrows and with a $U(1)$ gauge group.
\end{appendices}

\end{document}